\RequirePackage{hyperref}
\let\maketitle\relax
\documentclass{iucr}                      
\RequirePackage{graphicx}
\usepackage{todonotes}

\usepackage{tabularx}

\usepackage{graphicx}
\usepackage{dcolumn}
\usepackage{bm}

\usepackage{footmisc} 
\usepackage{breakurl}
\usepackage{ragged2e}

\usepackage{xcolor}

\newcommand{\widecaption}[1]{%
\refstepcounter{figure}%
\begin{justify}%
\small%
\textbf{\figurename\nobreakspace\thefigure\\}%
#1%
\end{justify}}
\usepackage{here}

\begin{document}                  



\title{Small-Angle X-ray Scattering: Characterization of cubic Au nanoparticles using Debye's scattering formula}


\cauthor[a]{Jérôme}{Deumer}{jerome.deumer@ptb.de}{address if different from \aff}
\cauthor[b]{Brian R.}{Pauw}{Brian.Pauw@bam.de}{address if different from \aff}
\cauthor[c]{Sylvie}{Marguet}{sylvie.marguet@cea.fr}{address if different from \aff}
\author[a]{Dieter}{Skroblin}
\author[c]{Olivier}{Taché}
\author[a]{Michael}{Krumrey}
\cauthor[a]{Christian}{Gollwitzer}{christian.gollwitzer@ptb.de}{address if different from \aff}

\aff[a]{Physikalisch-Technische Bundesanstalt (PTB), Abbestr. 2-12, 10587 Berlin, \country{Germany}}
\aff[b]{Federal Institute for Materials Research and Testing (BAM), Unter den Eichen 87, 12205 Berlin, \country{Germany}}
\aff[c]{Université Paris-Saclay, CEA, CNRS, NIMBE, 91191 Gif-sur-Yvette, \country{France}}









\maketitle                        


\begin{abstract}
We propose a versatile software package in the form of a Python extension, named \textbf{CDEF} (\textbf{C}omputing \textbf{D}ebye's scattering formula for \textbf{E}xtraordinary \textbf{F}ormfactors), to approximately calculate scattering profiles of arbitrarily shaped nanoparticles  for small-angle X-ray scattering (SAXS). CDEF generates a quasi-randomly distributed point cloud in the desired particle shape and then applies 
the open source software DEBYER for efficient evaluation of Debye's scattering formula to calculate the SAXS pattern  \cite{deumer2022}. If self-correlation of the scattering signal is not omitted, the quasi-random distribution provides faster convergence compared to a true random distribution of the scatterers, especially at higher momentum transfer.

The usage of the software is demonstrated for the evaluation of scattering data of Au nanocubes with rounded edges, which were measured at the four-crystal monochromator beamline of PTB at the synchrotron radiation facility BESSY II in Berlin. Our implementation is fast enough to run on a single desktop computer and perform model fits within minutes. The accuracy of the method was analyzed by comparison with analytically known form factors and verified with another implementation, the SPONGE, based on a similiar principle with fewer approximations. 
Additionally, the SPONGE coupled to McSAS3 allows us to further retrieve information on the uncertainty of the size distribution using a Monte-Carlo uncertainty estimation algorithm.
\end{abstract}


\section{\label{sec:Introduction}Introduction}
~\\[\baselineskip]
Small-angle X-ray scattering (SAXS) is a powerful nanostructure quantification tool to characterize ensembles of nanoparticles \cite{guinier_1955}. The X-ray scattering pattern of a nanoparticle system depends on many particle properties, which can therefore be obtained from the measurement, such as
the radius of gyration\cite{guinier_1955}, 
the particles' shape \cite{guinier_ellipsoid, guinier_1955, glatter_1982},
the size distribution \cite{size_dist_riseman}, 
specific surface area \cite{guinier_1955},
and number concentration \cite{schavkan_number_2019}. 

It is a non-destructive method with only little sample preparation for particles in liquid suspension, and also applicable for powders and porous materials \cite{porous_1997}. With SAXS, typically particles with sizes ranging from a few nanometers up to a few hundred nanometers can be measured, if there is sufficient electron density contrast of the particles relative to the suspension medium, since photons are scattered by the electrons in the material. The higher the electron density contrast, the more pronounced the scattered intensity relative to the background signal originating from the suspension. The measured SAXS signal can be further processed and fitted to gain information about the desired particle properties. 

To fit and evaluate experimental data, an adequate assumption of the underlying particle shape is necessary. This assumption is made by choosing the correct form factor $F(q)$ for the physical model, where $q$ is the magnitude of the photon's momentum transfer vector.

For simple particle shapes such as spheres, cylinders or spherical core-shell particles, $F(q)$ can be calculated analytically. For instance, $F(q)$ of a perfect sphere with a homogeneous electron contrast $\Delta \rho$ has been derived by \citeasnoun{rayleigh1911form}
\begin{equation}
F_{sph}(q, R, \Delta \rho) = \Delta \rho \, \Big( \frac{4}{3} \pi R^3 \Big) \, \Big(3 \frac{\sin qR - qR \cos qR}{(qR)^3} \Big), \label{eq:formfactor_sphere}
\end{equation}
where $R$ is the radius of the sphere and $q$ is the modulus of momentum transfer of the scattered photons.

The scattering pattern $I(q)$ of a polydisperse particle ensemble, as measured on the detector, is then obtained by convolving the absolute square of the form factor $\left|F_{sph}(q, R, \Delta \rho)\right|^2$ with the size distribution $g(R)$:
\begin{equation}
	I(q) = \int_0^\infty \left|F_{sph}(q, R, \Delta \rho)\right|^2 \cdot g(R) \mathrm{d}R \label{eq:intensity}
\end{equation}

Equation (\ref{eq:formfactor_sphere}) can be extended to other geometrical shapes with spherical symmetry such as core-shell particles, or particles with multiple concentric shells \cite{pedersen_2002, sasfit_manual}. For regular shapes with lower symmetry, the form factors are known, among many others, for ellipsoids \cite{guinier_ellipsoid},
cylinders \cite{guinier_1955}, cubic particles \cite{mittelbach1961rontgenkleinwinkelstreuung}, and for cylindrical and conical shaped particles with an arbitrary polygonal base which are built out of polygonal wedges \cite{shapovalov_light_2013}. For all these shapes, the average over all possible particle orientations is typically performed by numerical integration, and requires a one-dimensional average for shapes with one axis of rotational symmetry such as cylinders, and ellipsoids, and a two-dimensional for others like cubic shapes \cite{mittelbach1961rontgenkleinwinkelstreuung, formfaktor_cube_napper_1963, pedersen_2002, formfaktor_cube}, which is costly.

Recently, a seemingly limitless landscape of nanomaterial shapes and structures has been synthesized that do not fit these analytical functions such as stars \cite{zhou_anisotropic_2015, feld_chemistry_2019}, cubes with concave faces \cite{zhou_anisotropic_2015} or core-shell-structured cubes \cite{zhou_anisotropic_2015, jia2016thermosensitive, feld_chemistry_2019} demanding for a convenient method of calculating scattering profiles $I(q)$ of these complex shaped particles. Widespread SAXS analysis software such as \textit{SASfit} \cite{bresler_sasfit_2015}  or \textit{SasView} \cite{sasview} provide extended libraries of analytic form factors to evaluate SAXS data, however, analytic expressions for a particular shape may not be readily available and the derivation can quickly become intractable \cite{shapovalov_light_2013}.

An viable alternative approach to the analytic treatment of form factors for irregular shapes consists of building an approximation of the desired shape from smaller objects and calculating the scattering of the approximation via \possessivecite{debye_1915} scattering equation, which allows to directly compute the rotational average of an ensemble of scatterers from their individual form factors. \citeasnoun{hansen_1990} has proposed to build irregular shapes from randomly distributed point scatterers, and \citeasnoun{Pedersen_2012} successfully applied this method to the analysis of polydisperse ISCOM vaccine particles, which are perforated bilayer vesicles with or without proteins, composed of compounds with different scattering length densities.   

With the present paper, we introduce our open source software CDEF, which provides efficient calculation of approximate scattering profiles $I(q)$ for polydisperse ensembles of arbitrarily shaped nanoparticles. CDEF builds on the ideas put forward by \citeasnoun{hansen_1990} and \citeasnoun{Pedersen_2012} and enhances it with the option for quasi-random distribution of scatterers, which can improve convergence. An additional speed-up is achieved by offloading the actual calculation of the Debye formula to the open-source software DEBYER \cite{debyer}.

The algorithm is detailed in section \ref{sec:methods}. As an application, CDEF is used to evaluate scattering data from gold nanocubes with rounded edges in section \ref{sec:nanocubes}. Experimental details and the used nanomaterial are then described in section \ref{sec:exp} and \ref{sec:nanocubes}, respectively. Finally, the results are compared to the pre-existing program SPONGE based on similar principles \cite{aratsu_2020} in section \ref{sec:results}.

\columnbreak
\section{\label{sec:methods}Methods}
~\\[\baselineskip]
In this section, CDEF as well as the SPONGE will be described in more detail. Both programs are based on \possessivecite{debye_1915} scattering formula which can generally be used to calculate the SAXS pattern $I(q)$ of a system of $N$ individual scatterers 
\begin{eqnarray}
I(q) =& \sum_{k, j}^N f_kf_j \, \frac{\sin q r_{k,j}}{q r_{k,j}}
\label{eq:intensity2}
\end{eqnarray}
from the form factors $f_i$ of the individual scatterers and the distances $r_{k,j}$ between the scatterers $k$ and $j$.

\subsection{\label{sec:implementation_of_debyer}Implementation details of CDEF}

\begin{figure*}
\centering
\includegraphics[width=\textwidth]{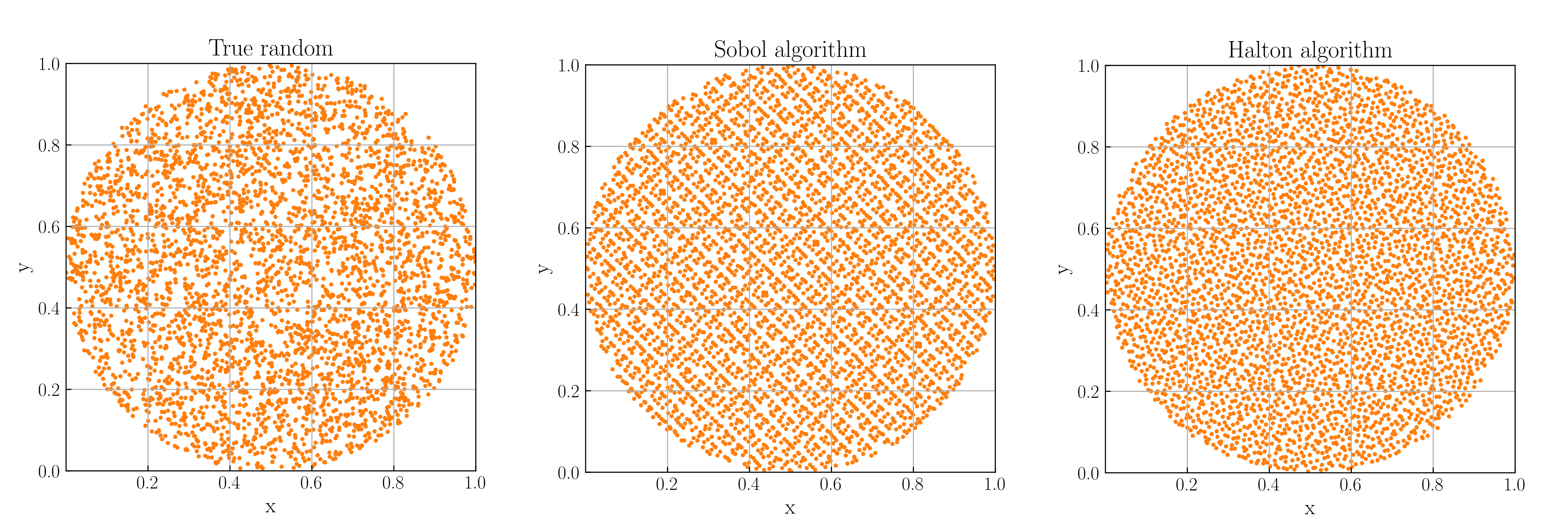}
\widecaption{Examples of circular point clouds with radius $r = 0.5$ generated with one true-random and two quasi-random (Sobol, Halton) filling algorithms. Each cloud is generated by initially filling $5000$ points into a squared area with side length $l = 1$ and subsequently deleting all points outside of the circle, which leaves $\sim 4000$ points. The usage of a quasi-random algorithm leads to a higher homogeneity of the spatial distribution relative to the true-random method whereas the true-random distribution shows a higher degree of local clustering.}
\label{fig:filling_algorithms}
\end{figure*}

To approximately calculate $I(q)$ for arbitrarily shaped nanoparticles, CDEF applies the equation \ref{eq:intensity2} onto a three dimensional point cloud of the desired particle shape. The point cloud is created by filling the particle's bounding box with equally distributed punctiform scatterers and discarding all points outside of the volume defined by the particle's shape. The shape can be built either from a CAD construction, for this both CDEF and the SPONGE offer the import of the widely used STL file format, or by programmatically reshaping the point cloud.
CDEF provides the option to generate the intial point cloud from either a true- or quasi-random sequence (fig. \ref{fig:filling_algorithms}). Compared to the true random series, a quasi-random sequence fills the shape more evenly with less local clustering \cite{bart_good_2006}. We implemented a generator for the scrambled \citeasnoun{halton1964algorithm} series proposed by \citeasnoun{kocis1997computational} and \possessivecite{sobol_1967} series as provided by the SciPy package \cite{SciPyNMeth2020}.

Each point of the generated cloud gets then assigned a weight to account for density variations such as in heterogeneous or core-shell particles. 
Finally, as a computational tool to efficiently evaluate Debye's scattering formula, CDEF passes the points and the associated weights to the open-source program DEBYER \cite{debyer}, whereas the SPONGE uses its own implementation of the Debye equation.
Similar approaches to compute form factors for arbitrary shapes using Debye's scattering formula have been mentioned by \citeasnoun{pedersen_2002}, \citeasnoun{Pedersen_2012}, or \citeasnoun{hansen_1990}, and are used by other fast programs, e.g., DEBUSSY \cite{debussy_2015}.

A more detailed comparison between CDEF, SPONGE and other evaluation methods using Debye's equation can be found in the supporting information (SI).

As proposed by \citeasnoun{hansen_1990} and \citeasnoun{Pedersen_2012}, DEBYER achieves a significant performance gain by splitting the calculation of equation \ref{eq:intensity2} into two parts.
First a histogram of the pair distances $r_{k,j}$ is computed with a reduced number of histogram bins $N_{\mathrm{BINS}}$ and subsequently the sinc function $ \frac{\sin q r}{q r}$ is evaluated for each bin of the histogram. Because $N_{\mathrm{BINS}}$ is usually much smaller, typically around $1,000$ to $10,000$, than the number of pairs of scatterers $N^2$, this approximation can speed up the computation by several orders of magnitude for repeated evaluation of equation \ref{eq:intensity2} for different $q$, such as in the computation of a full scattering pattern. 
CDEF allows to set the histogram bin width explicitly to trade off the accuracy of the computed scattering curve with computation time.

The scattering pattern $I_{\mathrm{MONO}}$ obtained in this way corresponds to a single particle, averaged over all possible orientations. For the modelling of realistic particle dispersions, $I_{\mathrm{MONO}}$ must be averaged over a certain size distribution. CDEF achieves this by rescaling the single-particle scattering curve from a single master curve according to
\begin{eqnarray}
	I_{\mathrm{POLY}}(q)  = \int_0^\infty V^2 I_{\mathrm{MONO}}(qR) \cdot g(R) \mathrm{d}R,
\label{eq:intensity8}
\end{eqnarray}
which avoids repeated evaluation of Debye's scattering equation for different particle sizes. Here, $g(R)$ is the size distribution and $V$ the volume of the rescaled particle with size $R$.
The integral in equation \ref{eq:intensity8} is evaluated by Monte Carlo integration with $3000$ samples using a normal random number generator, yielding a Gaussian size distribution, but other distributions can be easily implemented by using the appropriate random number generator. At the moment, CDEF implements Gaussian and lognormal distributions.

The implementation of a Poisson Disc algorithm to fill the bounding box homogeneously with scatterers which are required to have a certain minimum distance to each other would also be conceivable. However, this requires more computational effort, e.g. filling a cube with $30 \, 000$ points is approximately $28$ times slower ($\sim 350 \, \mathrm{ms}$ (Sobol) vs. $9.85 \, \mathrm{s}$), and would not bring any apparent advantages over the existing algorithms (SI).

\subsection{\label{sec:DEBYER_vs_ana}CDEF vs. analytic formulae}
\begin{figure}
\includegraphics[width=\linewidth]{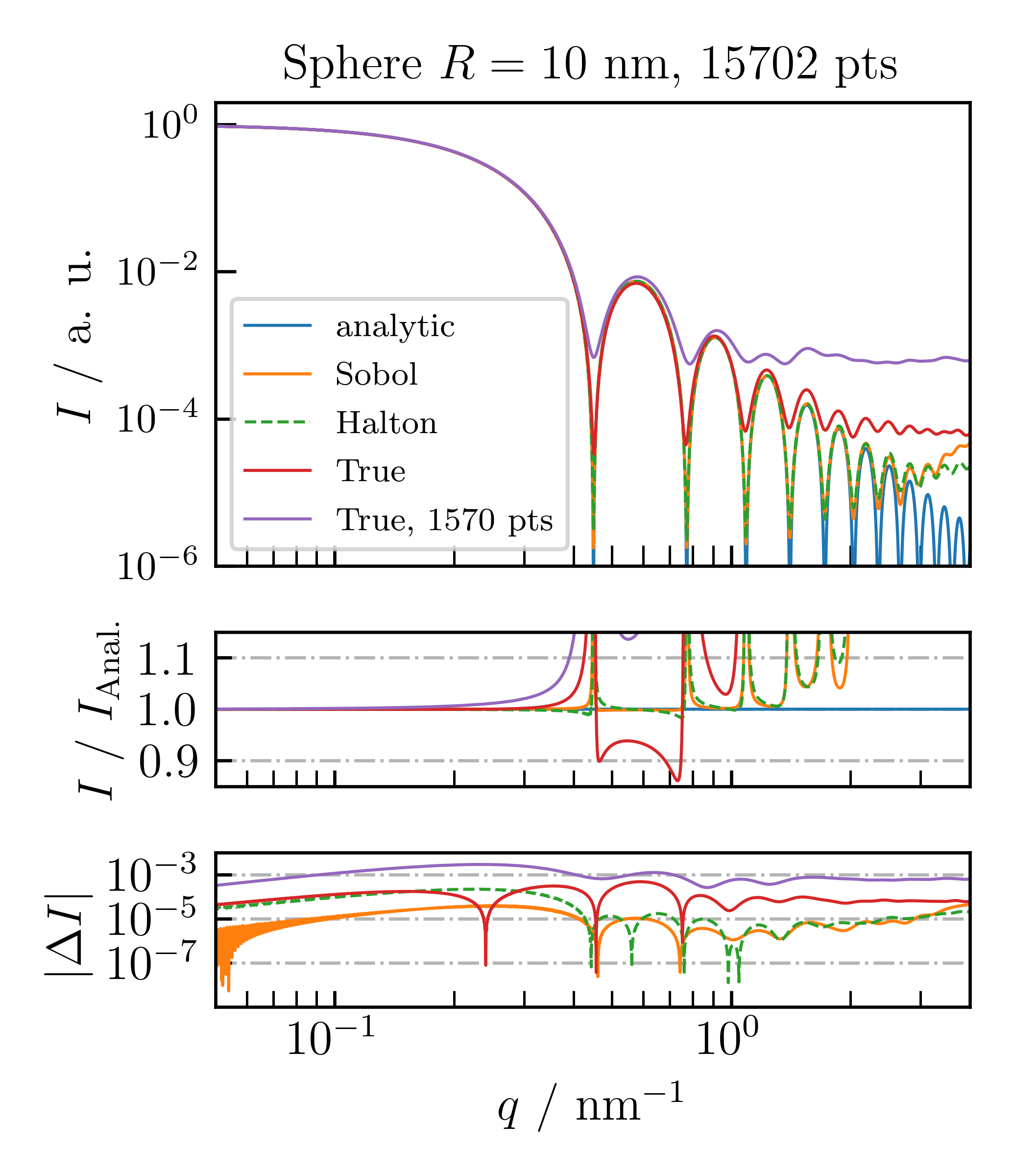}
\caption{Comparison of normalized single-particle SAXS profiles using CDEF \textit{without} modeling of the artificial background signal with the exact analytic SAXS profile $I_{\mathrm{Anal.}}$ of a sphere with radius $R=10 \, \mathrm{nm}$ and electron contrast $\Delta \rho = 1 \, \mathrm{nm}^{-3}$. For the numeric calculations, the Deybe equation was applied on spherical clouds which were generated using two different quasi-random (Sobol, Halton) and one true-random filling algorithms. At specific $q$-values the artificial scattering signal from the fine structure of the individual cloud dominates the numeric profiles leading to a deviation from $I_{\mathrm{Anal.}}$.}
\label{fig:debyer_result_1}
\end{figure}

As a validation of CDEF, we first compare its normalized results with the corresponding analytic form factors of common particle shapes using the three introduced filling algorithms (fig. \ref{fig:filling_algorithms}). Fig. \ref{fig:debyer_result_1} shows the analytically (eq. \ref{eq:formfactor_sphere}) and numerically calculated single-particle SAXS profiles of a homogeneous sphere with radius $R = 10 \, \mathrm{nm}$. For the calculation of each numeric profile, a spherical cloud was generated by (quasi-)randomly filling $30 \, 000$ points into a cubic bounding box with side length $2R = 20 \, \mathrm{nm}$ and then deleting all points outside of the defined sphere, which yields approximately $N \approx 15 \, 700$ remaining points. 

Both quasi-random profiles match the analytic profile in good agreement up to the $5$th local maximum, whereas at higher $q$-values both profiles start deviating from the analytic profile due to an artificial background signal originating from the clouds' fine structure. This also holds true for the true-random filling pattern with the same number of scattering points, however, it only matches $I_{\mathrm{Anal.}}$ up to the $2$th local maximum due to a constant scattering background. Fig. \ref{fig:debyer_result_1} also shows that reducing the number of scattering points by a factor of 10 raises the background plateau by the same factor.

\citeasnoun{Pedersen_2012} proposed that the constant background arising from the true random distribution can be subtracted by excluding the self-correlation of the scatterers, which corresponds to zeroing the first bin of the pair distance histogram or subtraction of a constant value of $1 \, \mathrm{/} \, \mathrm{N}$ from the resulting scattering patterns. This does indeed increase the dynamic range of the computed scattering curve and brings it in closer agreement with the true pattern. 

\begin{figure}
\includegraphics[width=\linewidth]{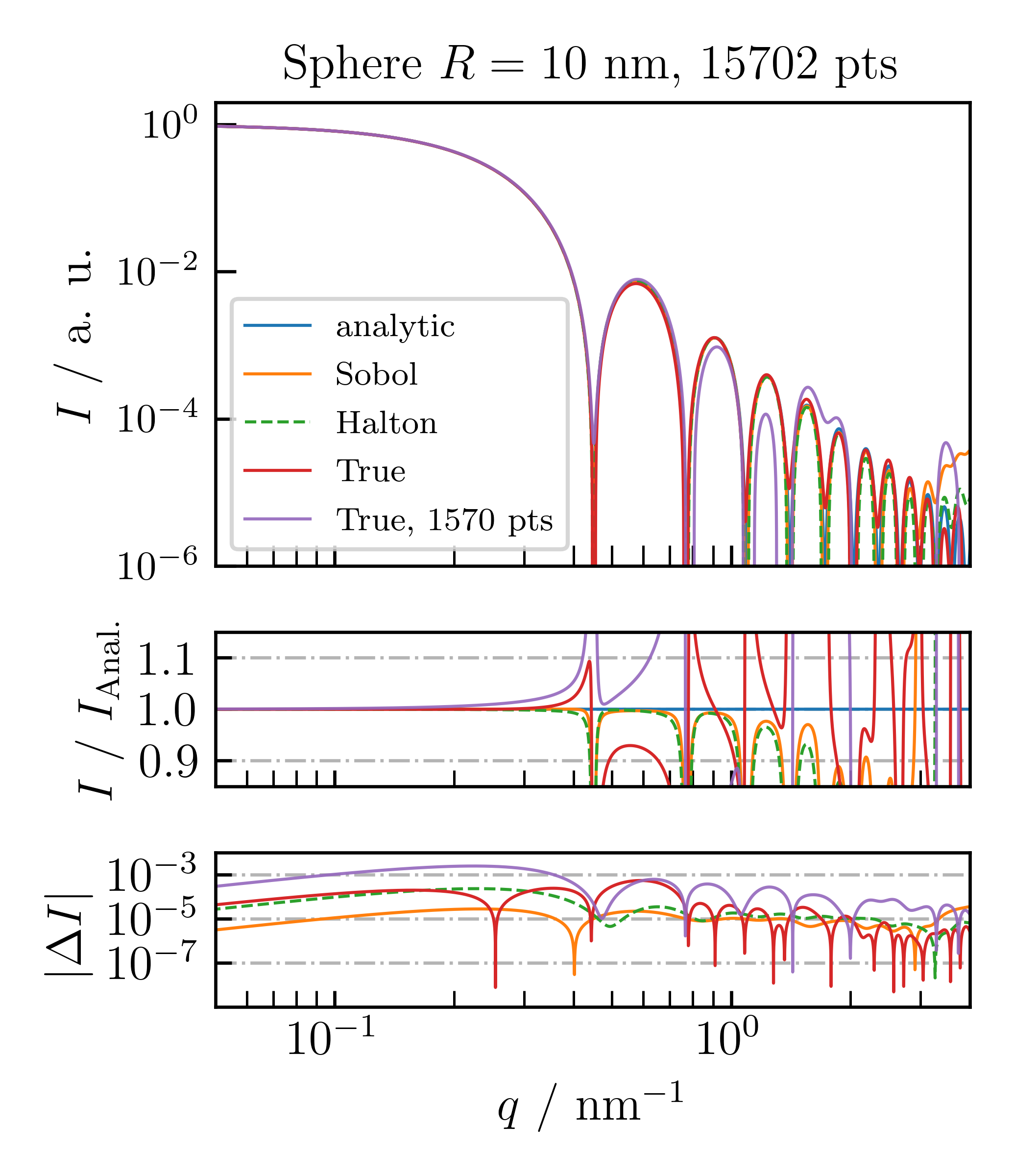}
\caption{Comparison of normalized single-particle SAXS profiles with modeling of the artificial background signal.} 
\label{fig:debyer_result_1_1}
\end{figure}

\columnbreak
Fig. \ref{fig:debyer_result_1_1} displays background-corrected scattering patterns for the three different filling algorithms. For the quasi-random filling algorithms, zeroing the first bin does not improve the agreement with the exact solution because of the low autocorrelation at small distances of quasi-random sequences. Instead, zeroing a small initial sequence of bins except for the first can bring curves in closer agreement with the exact result (see fig.  \ref{fig:debyer_result_1_1}). Still, the curve computed from the quasi-random sequences without this correction is in better agreement for midrange values of $q$ than the corrected true random solution, which is evident by comparing the plots with the relative deviation in figs. \ref{fig:debyer_result_1} and \ref{fig:debyer_result_1_1}. To perform these optimizations for a given case, CDEF provides the option to zero out a sequence of bins in the pair distance distribution histogram.

Similar results are obtained for a comparison of particles with lower symmetry, such as cylinders and cubes. The corresponding data can be found in the supporting information (figs. S$2$, S$5$, S$6$, S$7$).

\subsection{\label{sec:sponge} The SPONGE}

\subsubsection{\label{sec:implementation_of_sponge} Implementation details of the SPONGE}
A separate implementation was developed, called the SPONGE \cite{sponge_git}, that is a more fundamentally proximate method by eschewing many of the speed-improving approximations. It also uses the Debye equation for puctiform scatterers with a fully random point distribution. This method is essentially similar to CDEF with the exception that the intermediate step where the numerical pair distance distribution function is generated, is bypassed in favor of a more direct approach, further minimizing potential sources of error. 

While the SPONGE is much more computationally intensive, it should be more accurate over the entire $q$-range where the homogeneous phase approximation holds, and thus can be used to validate that the approximations in the faster CDEF implementations are not generating unforeseen artefacts. Like CDEF, the SPONGE uses the surface description in the STL format to define the boundaries of a nano-object. It then leverages the fast VTK bindings in Python \cite{VTK_2006} for point placement, and determination whether the point lies inside or outside of the object. The computation of the point-to-point Euclidian distance matrix is done using a fast SciPy implementation \cite{scipy_distance}, before the Debye equation is applied to obtain a simulated isotropic scattering curve. By entering a scattering length density, the SPONGE-simulated data can be scaled to absolute units (i.e. to an absolute scattering cross-section in $\mathrm{m}^{-1} \, \mathrm{sr}^{-1}$).

This procedure is then repeated, resulting in a number of independently generated scattering curves, each based on their own set of random points. The mean intensity from all repetitions is then presented, with the standard deviation used as an estimate for the uncertainty for each point. 

A number-weighted size distribution can also be taken into account. The SPONGE currently uses a Gaussian size distribution, which is implemented by picking a random scaling factor for the $q$-vector for each independent repetition, and which affects the total intensity scaling factor based on its scaled volume (in a procedure identical to that as given in section \ref{sec:implementation_of_debyer}). This would be similar in reality to probing a multitude of objects of different size to build up the average scattering pattern. This size distribution has been verified to work accurately (by checking the result with a fit in SasView) up to a Gaussian distribution width $\sigma$ of at least $50\,\%$. This simulated distribution width is not used for fitting, but is used to avoid unrealistically sharp minima in the simulated curve. For the simulations presented herein, the distribution width is set to $1\,\%$.

\subsubsection{\label{sec:sponge_und_mcsas} The SPONGE and MCSAS3}

The thus simulated data of primary particles can be used to fit an experimental dataset, even when the experimental dataset is from a sample with an unknown, broader distribution of particle sizes. For this, we turn the simulated data into a fitting model for use with the Monte Carlo approach as implemented in McSAS \cite{MCSAS_2013, MCSAS_2015}. As the original McSAS is not easily adapted to support such a model description, we are here using the refactored McSAS3 implementation (currently in the last stages of development). McSAS3 works using the same methods as McSAS, but has many practical improvements such as multi-threaded optimization, a gui-independent backend (for headless computation), and the option to re-histogram a previous optimization run (McSAS on GitHub \cite{mcsas_git}).

The simulated data can be converted into a fitting model, provided it has a Guinier region at low $q$, and (on average) a Porod region at high-$q$. Then, for a given scaling factor, the $q$ vector of the simulated data is rescaled (in a manner identical to section III), and the intensity interpolated to the requested $q$ vector of the experimental data. Datapoints that fall outside the limits of the simulated data are extrapolated using a flat (Guinier) approximation at low $q$, and a Porod slope at high $q$.

Using this fitting model in McSAS3, experimental data can be fitted rapidly using the simulated scattering pattern of an elementary scatterer. From this, a form-free volume-weighted scaling factor distribution is obtained that best describes the experimental data. As with the original McSAS \cite{MCSAS_2013}, a number of independent optimizations are performed to allow the estimation of the uncertainty of the resulting distribution.

\subsection{Diverse models of cubic particles}

\begin{figure}
\includegraphics[width=\linewidth]{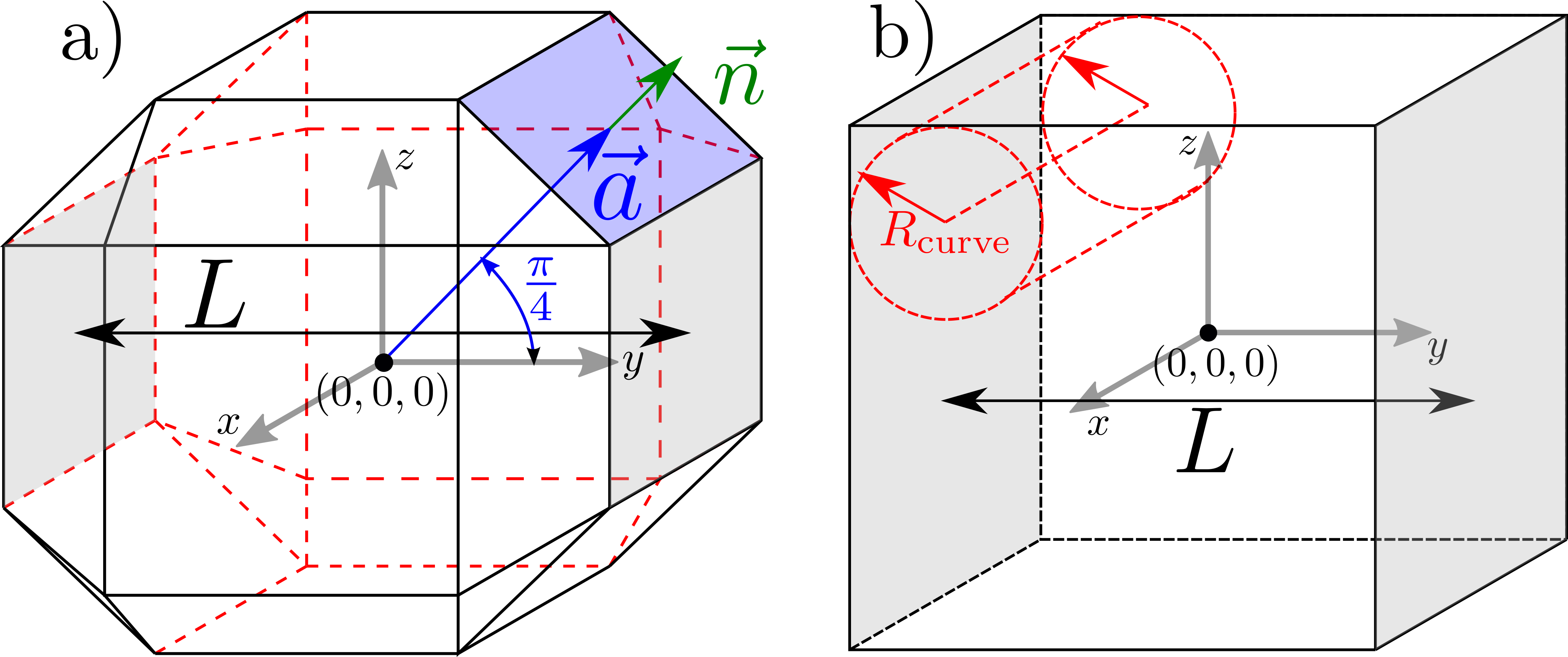}
\caption{Two different cubic models with face-to-face-distance $L$. a) Cube with truncated edges. All 12 edges are truncated by sectional planes. Each sectional plane, marked blue, is defined by a support vector $\vec{a}$, and a normal vector $\vec{n}$ which stands perpendicular on the plane. b) Cube with implied rounded edges. The curve of each edge is defined by identical cylinders with curvature radius $R_{\mathrm{curve}}$ which touch the associated cubic sides.}
\label{fig:wuerfel_mit_kanten}
\end{figure}

To show the versatile application of CDEF, we want to characterize Au nanocubes which are described in section \ref{sec:nanocubes}. In doing so, we implemented three different cubic models (ideal cube, cube with truncated edges, cube with rounded edges) carrying a homogeneous electron density (fig. \ref{fig:wuerfel_mit_kanten}). 

To simulate truncated edges, an advanced (i.e. point clouds are generated by user-written python functions) algorithm based on the Hessian normal form is implemented with which the truncation-level of the cubic model with a side-to-side distance $L$ can be adjusted. Further information will be provided in the supporting information. 

Moreover, a cubic model with rounded edges is generated by introducing four cylinders for each Euclidean direction $x$, $y$, $z$ whereas each cylinder is located in one of the four corners with its axis being aligned along the corresponding edge (fig. \ref{fig:wuerfel_mit_kanten}). The rounded edges are now generated by deleting points, i.e. setting their corresponding form factor to zero, located at the edges and outside of each cylinder. All 12 cylinders are described by the same radius of curvature $R_{\mathrm{curve}}$.

\section{\label{sec:nanocubes}Synthesis of Au nanocubes}
\begin{figure}
\includegraphics[width=\linewidth]{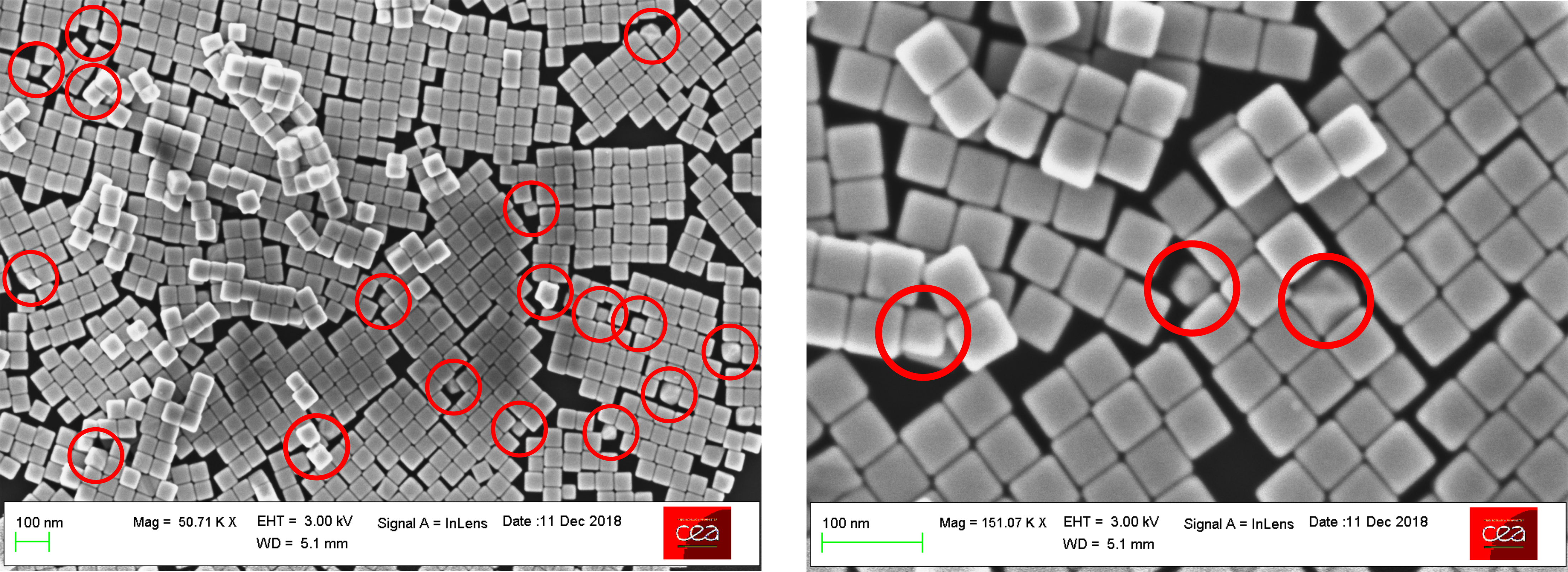}
\caption{Scanning electron microscopy images of Au nanocubes at two different scales. The population also consists of particles with a non-cubic shape (some marked by red circles).}
\label{fig:SEM_nanocubes}
\end{figure}

~\\[\baselineskip]
Mono-crystalline Au nanocubes  (fig. \ref{fig:SEM_nanocubes}) were prepared by colloidal chemistry in aqueous solution, according to an already published protocol \cite{cube1, cubes_ref_2}, in presence of cetyltrimethylammonium bromide (CTAB) as the capping agent. Crystal growth is achieved by chemical reduction of Au+ ions on the surface of a gold seed (a small sphere with an initial size of $2$ to $3$ nm in diameter), resulting in the formation of a cubic shape \cite{cubes_ref_3}. The side length of these particles as determined from SEM images amounts to $55\,\mathrm{nm}$ with a standard deviation of $2\,\mathrm{nm}$. Using this particular synthesis procedure leads to a percentage of $\sim 90\,\% $ of nanocubes w. r. t. the whole particle ensemble and a small amount $\sim 10\,\%$ of particles with different shapes (see marked spots in fig. \ref{fig:SEM_nanocubes}). The edges and corners of the cubes tend to gradually round out over time. In solution, this phenomenon is slow (6 months), however, it is faster (1 month) when the cubes are deposited on a substrate and kept in air. From the SEM images, a curvature radius ($R_{\mathrm{curve}} \approx 7 \, \mathrm{nm}$) was determined for the edges.

\section{\label{sec:exp}Experimental details}
~\\[\baselineskip]
Since the SAXS experiments were conducted in vacuum, the diluted colloidal solution of Au nanocubes suspended in water was filled into a rectangular capillary of borosilicate glas with a homogeneous thickness along its vertical axis, and sealed with a blow torch before measurement. 
The sample was then loaded into the experimental vacuum chamber which is connected to the four-crystal monochromator (FCM) beamline of the PTB laboratory at the synchrotron radiation facility BESSY II, Berlin. For the experiment, X-rays are generated by a bending magnet, and then guided by the beamline to the sample holder resulting in a thin X-ray beam with a cross-sectional area of approximately $150 \, \mu \mathrm{m} \, \times \, 400 \, \mu \mathrm{m}$ at the samples' position.
The FCM-beamline allows to perform experiments in a wide range of photon energies from $E_{\mathrm{ph}} = 1.75 \, \mathrm{keV}$ to $10 \, \mathrm{keV}$ \cite{krumrey_design_1998}. Our SAXS experiments were performed at $E_{\mathrm{ph}} = 8 \, \mathrm{keV}$ using the Si $(1,1,1)$ monochromator crystals with a spectral resolving power of $E_{\mathrm{ph}} \, \mathrm{/} \, \Delta E_{\mathrm{ph}} = 10^4$ and a photon flux in the range of $\Phi \approx 10^{10} \mathrm{/} \mathrm{s}$ \cite{krumrey_design_1998}. During the experiment, the capillaries were measured at different $y$-positions along the vertical axis. At each $y$-position, SAXS images were recorded by a vacuum-compatible PILATUS 1M hybrid-pixel detector with a pixel size of $p = 172 \, \mu \mathrm{m}$ \cite{wernecke_characterization_2014}.

\subsection{\label{sec:processing}Data processing}

Prior to data evaluation, the two dimensional (2D) SAXS image, consisting of concentric circles, is converted into the corresponding one dimensional SAXS profile in absolute units. This allows us to determine the number concentration of suspended particles. For each distinct $y$-position, $I_{\mathrm{EXP}}$ is circularly integrated around the center of the incident beam and then normalized to the incident photon flux, the duration of exposure, the sample thickness and the quantum efficiency of the detector at a given photon energy \cite{schavkan_number_2019}. Then $I_{\mathrm{EXP}}$ is expressed in terms of the momentum transfer $q$:

\begin{eqnarray}
q = \frac{4 \pi E_{\mathrm{ph}}}{h c} \sin \Big( \frac{1}{2} \arctan \frac{n p}{L_{\mathrm{SD}}} \Big) \approx \frac{4 \pi E_{\mathrm{ph}}}{h c} \frac{n p}{L_{\mathrm{SD}}},
\label{eq:q}
\end{eqnarray}
where $L_{\mathrm{SD}}$ is the distance from the sample to the detector plane, $n$ is the number of pixels, and $p$ is the pixel size. Data processing at PTB, up to this point, is standardized using particular in-house software. 

Since scattering from water molecules and the walls of the glass capillary is also detected by the SAXS measurement leading to an unwanted background signal, an additional capillary only filled with distilled water was measured during the same measurement to detect the corresponding background curve by which $I_{\mathrm{EXP}}$ was eventually subtracted. For better statistics, however, $I_{\mathrm{EXP}}$ as well as the background curves were averaged over all $y$-positions before subtraction.

After subtraction of the background signal, it was not necessary to include an independent background in the fitting model. This also reduces the number of adjustable parameters.

\section{\label{sec:results}Results and Discussion}
~\\[\baselineskip]
In this work, we characterized Au nanocubes using three different cubic models, namely an \textit{ideal} cube, a cube with truncated edges as well as a cube with rounded edges. However, for reasons of convenience, only results referring to the model with rounded edges, which shows the lowest $\chi^2$ (tab. \ref{tab:table1}), are presented. Detailed results of the other models can be found in the supporting information (SI). 

During the fitting of shapes with varying geometry, such as the truncated or rounded cubic model, it is necessary to recalculate the individual single-particle scattering profile $I_{\mathrm{MONO}}$ in each computational step. For steady particle shapes with size changes only, such as ideal cubes, it is sufficient to calculate $I_{\mathrm{MONO}}$ once and then rescale it in accordance to the assumed size distribution which requires much less computational effort. For all models, no modeling of the artificial background signal was performed as illustrated in section \ref{sec:DEBYER_vs_ana} whereas a sufficient high number of $N = 30 \, 000$ scatterers was chosen to cover the required $q$-range.

\begin{figure}
\includegraphics{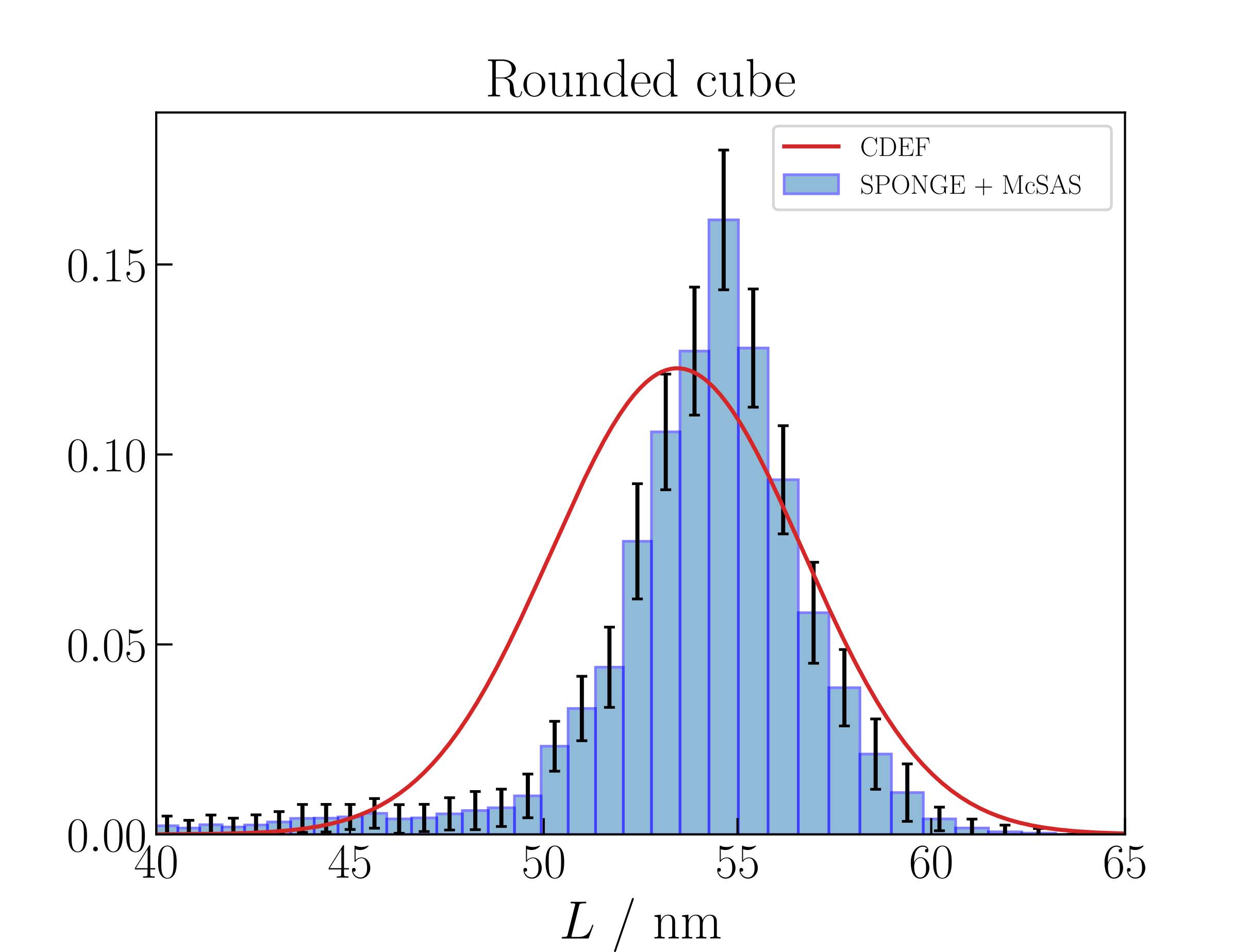}
\caption{CDEF vs. the SPONGE: the SPONGE's volume-weighted size distribution reveals a mean value of $L = (54.00 \pm 0.06) \, \mathrm{nm}$. The volume-weighted distribution using CDEF again shows an expectation value of $L = 53.4 \, \mathrm{nm}$.}
\label{fig:hist_rounded}
\end{figure}

For all introduced cubic models, the results of the (faster) CDEF are compared to those of the SPONGE to confirm the results of CDEF. Figure 
\ref{fig:hist_rounded} compares the volume-weighted size distribution from the SPONGE with the size distribution from CDEF, converted into volume weight. In doing so, both methods evaluated the same experimental data. Since the SPONGE cannot fit shape parameters due to the time-taking computing process, stl-files of CDEF's best-fit particle shapes were generated and then given to the SPONGE to reveal the underlying uncertainty of $I_{\mathrm{FIT}}$.

\begin{figure*}
\centering
\includegraphics[width=1.\textwidth]{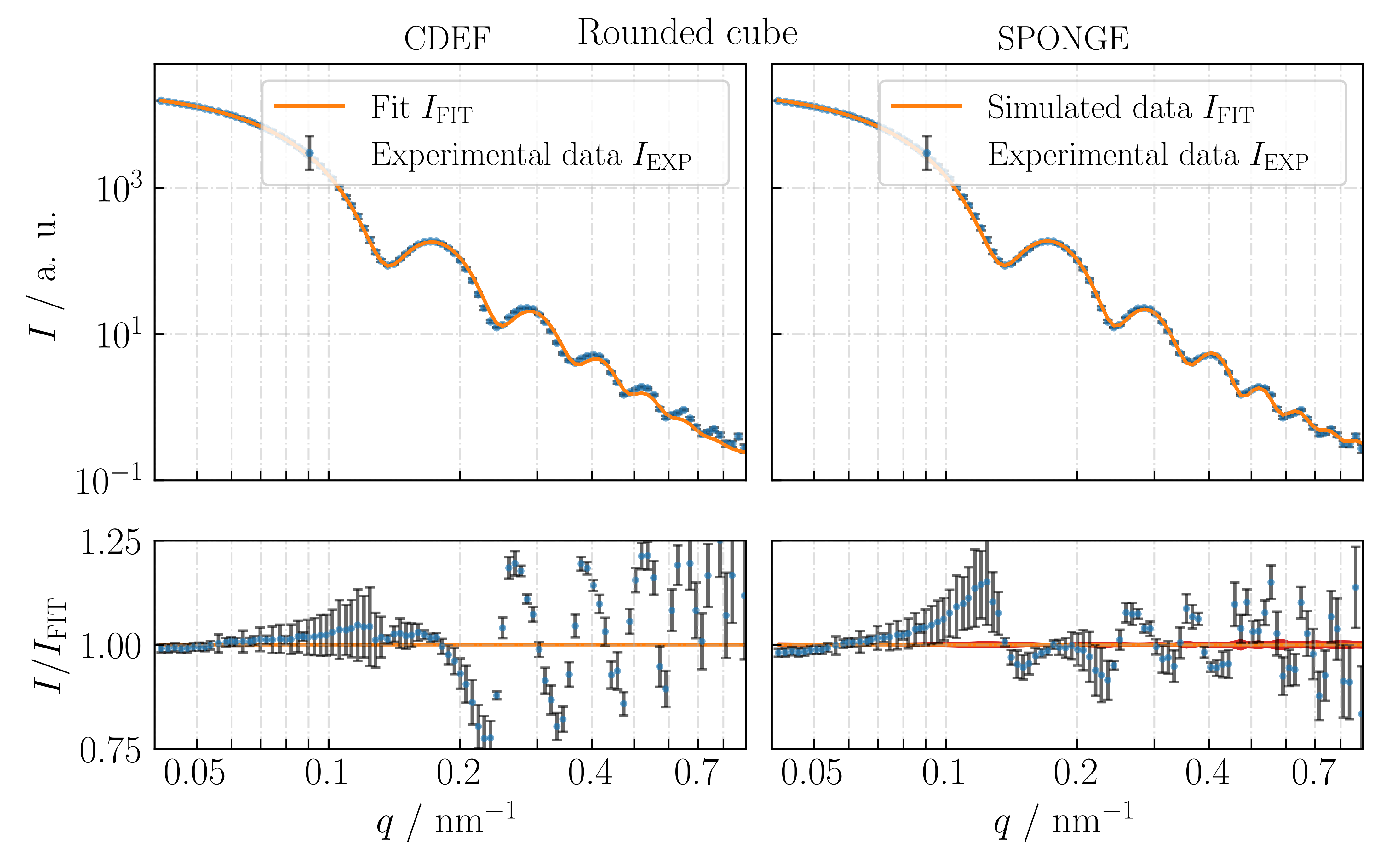}
\widecaption{CDEF versus the SPONGE. Fit results of Au nanocubes using a cubic model with rounded edges. Coupling of the SPONGE with MCSAS additionally reveals an uncertainty of $I_{\mathrm{FIT}}$, thus an uncertainty of the underlying size distribution can be stated (fig. \ref{fig:hist_rounded}). See table \ref{tab:table1} for further information.}
\label{fig:saxs_rounded} 
\end{figure*}

Using CDEF, each model was fitted to the experimental data by varying the $M$ free parameters, namely the number-weighted distribution of the side-to-side length $L$, which was assumed to be Gaussian, and the truncation or rounding parameters for the imperfect cubes. Powell's algorithm \cite{powell_1963} with a maximal number of $M \cdot 1000$ function evaluations was used to minimize $\chi^2$. 
The combined SPONGE$+$McSAS was not confined to any particular size distribution, but rather fitted the volume-weighted size distribution numerically.

With CDEF, each 3D cloud initially consisted of $N=30 \, 000$ scattering points whereas for each function evaluation step $N$ varied based on the underlying spatial distribution of scatterers, the level of truncation $T$ or radius of curvature $R_{\mathrm{curve}}$ such that $N < 30 \, 000$. This initial number of $N=30 \, 000$ was a good compromise to fit the whole $q$-interval of the experimental data without experiencing any artifacts arising from the clouds' fine structure, but staying below a computing duration of $< 4 \, \mathrm{s}$ per evaluation of $\chi^2$. For comparison, a spherical model was additionally included in the evaluation (tab. \ref{tab:table1} \& SI).

For both methods, the cubic models with truncated (SI) or rounded edges (fig. \ref{fig:saxs_rounded}) fit the experimental data slightly better than the ideal cube. The lower values of $\chi^2$ (table \ref{tab:table1}) also imply a higher degree of compliance for these models which coincides with the fact that the particles' edges and corner gradually round out over time when being stored in suspension above $6$ months.

For the model with rounded edges we obtain the same result of $L = 53.4 \, \mathrm{nm}$ with $\sigma_{\mathrm{L}} = 3.2 \, \mathrm{nm}$. With this model, we additionally obtain a radius of curvature of $R_{\mathrm{curve}} \approx 7 \, \mathrm{nm}$ which is in good agreement with the value measured with SEM (section \ref{sec:nanocubes}).  With the SPONGE we obtain $L = (54.00 \pm 0.06) \, \mathrm{nm}$ and $\sigma_{\mathrm{L}} = (3.1 \pm 0.9) \, \mathrm{nm}$. The relative deviation of the mean face-to-face-distance $\Delta L \, \mathrm{/} \, L$ again equals $1.1\,\%$.

\begin{table*}
\caption*{\label{tab:table1}CDEF: Summary of fitting results of homogeneous cubic models with number concentration $C$, mean particle size $L$, standard deviation $\sigma_{L}$, truncation factor $T$ (in terms of $L \mathrm{/} \sqrt{2}$), radius of curvature $R_{\mathrm{curve}}$ (in terms of  $L \mathrm{/} 2$), number of iterations $N_{\mathrm{iter}}$ of Powell algorithm, number of function evaluations $N_{\mathrm{fev}}$ of function $ \chi^2$ and computing time $t$.}

\begin{tabular*}{\linewidth}{l|lllllllll}
Model & $C$ \footnote{Number concentration is based on an electron contrast of $\Delta \rho \approx 4077 \, \mathrm{nm}^{-3}$ of Au-particles suspended in $\mathrm{H}_2 \mathrm{O}$ at $8 \, \mathrm{keV}$.}  / $\mathrm{cm}^{-3}$ & $L$ / nm & $\sigma_{\mathrm{L}}$ / nm & $T$ / 1 & $R_{\mathrm{curve}}$ / 1 & $ \chi^2$ & $N_{\mathrm{iter}}$ & $N_{\mathrm{fev}}$ & $t$ / s \\ \hline
\\[-0.5em]

Ideal cube & $8.709 \cdot 10^{9}$ &$52.5$   & $2.8$  & & & $<33$ & $5$ & $262$ & $< 37$ \\

Truncated cube & $8.604 \cdot 10^{9}$ &$53.4$ & $3.3$  & $0.91$ & & $<23$ & $5$ & $433$ & $< 1339$ \\

Rounded cube & $8.562 \cdot 10^{9}$ &$53.4$  & $3.3$  & & $0.27$ ($7.2  \, \mathrm{nm}$) & $< 21$ & $6$ & $493$ & $< 1172$ \\

Sphere & $8.636 \cdot 10^{9}$ &$31.7$\footnote{Spherical radius in nm.} &  $3.4$\footnote{Standard deviation of radius in nm.}  & & & $< 158$ & $5$ & $274$ & $< 39$
\end{tabular*}
\end{table*}

Since the measured ensemble of Au nanocubes does not only consist of cubes with a single shape (ideal, truncated, or rounded), but partially contains all of them plus particles with undefined (i.e. non-cubic) shapes (fig. \ref{fig:SEM_nanocubes}), none of the specific cubic models used is actually able to exactly fit the measurement data, meaning $I \mathrm{/} I_{\mathrm{FIT}} \approx 1$ for the entire $q$ range, respectively $\chi^2 \le 1$. Also the uncertainty estimate coming from data processing (chapter \ref{sec:processing}), meaning the background subtraction in particular, could be underestimated.  

Thus, a next step to improve the overall model of the particle ensemble could be the application of a model function including all assumed cubic models with their volume-weighted percentage of the total particle population. The percentage would need to be determined for a representative sample of the ensemble in advance, for instance using microscopic methods with which number-weighted percentages would be obtained.

\section{Conclusion}
~\\[\baselineskip]
CDEF is suitable to calculate single-particle SAXS profiles of common particle shapes (including shapes with high aspect ratios) with satisfying accuracy which was shown by comparison with known analytic form factors. Here, a sufficient but minimal number of scattering points should be selected though to prevent artifacts from appearing in the scattering profile but keeping computing effort low. Additionally, user of CDEF are able to make manual changes to the underlying pair distance histogram to further reduce the number of necessary scattering points. 
Occasional cross-checks can be made between CDEF and the SPONGE to ensure the speed improving assumptions in CDEF are not interfering with the accuracy of the results. Using CDEF, polydisperse SAXS pattern can also be generated, eventually allowing experimental data to be evaluated. For all presented cubic models, a direct comparison between CDEF and the SPONGE concerning the size distribution of Au nanocubes reveals good agreement between results, with a deviation of the mean size of $\le 1.5\,\%$, even though CDEF uses the histogram approximation of the pair distances through DEBYER and is confined to a Gaussian distribution. 
The time-saving approach of implementing Debye’s equation in CDEF further allows to introduce fit parameters of the particle shape which enable users to gain more detailed information of the measured nanoparticles. In terms of "steady-shape" particle evaluation, moreover, CDEF has also been coupled with a Markov-Chain-Monte-Carlo algorithm to additionally reveal uncertainty estimates of the assumed size distribution of bipyramidal Ti$\mathrm{O}_2$ nanoparticles \cite{loic_2021}. 

While the more direct SPONGE approach is not quick enough for iterative optimization methods, the coupling of the SPONGE with McSAS3 allows the determination of size distributions of odd-shaped particles when no information on the shape of the analytical size distribution is known. 
The coupling of CDEF with McSAS3 is, in principle, also possible since both programs are implemented as Python libraries. This would lead to superior performance compared to SPONGE and will be considered in future versions of CDEF. 

Both approaches can be extended to include core-shell morphologies by varying the density of scatterers or assigning different electron densities to the individual punctiform scatterers. Further speed-up could be achieved by an implementation which runs on parallel hardware such as consumer graphics cards.










\section{Acknowledgements}
\ack{This work was partly funded by the 17NRM04 nPSize project of the EMPIR programme co-financed by the EMPIR participating states and by the European Union's Horizon 2020 research and innovation programme. 

Finally, we want to thank Levent Cibik and Jan Weser from PTB for their help constructing different particle shapes using CAD as well as for their beamline support.}



\referencelist{iucr}





\end{document}





\title{Small-Angle X-ray Scattering: Characterization of cubic Au nanoparticles using Debye's scattering formula \\[1ex] \large Supporting Information}


\cauthor[a]{Jérôme}{Deumer}{jerome.deumer@ptb.de}{address if different from \aff}
\cauthor[b]{Brian R.}{Pauw}{Brian.Pauw@bam.de}{address if different from \aff}
\cauthor[c]{Sylvie}{Marguet}{sylvie.marguet@cea.fr}{address if different from \aff}
\author[a]{Dieter}{Skroblin}
\author[c]{Olivier}{Taché}
\author[a]{Michael}{Krumrey}
\cauthor[a]{Christian}{Gollwitzer}{christian.gollwitzer@ptb.de}{address if different from \aff}

\aff[a]{Physikalisch-Technische Bundesanstalt (PTB), Abbestr. 2-12, 10587 Berlin, \country{Germany}}
\aff[b]{Federal Institute for Materials Research and Testing (BAM), Unter den Eichen 87, 12205 Berlin, \country{Germany}}
\aff[c]{Université Paris-Saclay, CEA, CNRS, NIMBE, 91191 Gif-sur-Yvette, \country{France}}









\maketitle                        


This is the Supplementary Information (SI) to the paper describing CDEF, a software library to compute scattering form factors for arbitrarily shaped particles. Among other information, the SI contains additional simulated and fitted scattering curves. The chapters in this SI correspond directly to the chapters with the same number in the main manuscript.


\newpage
\section{Implementation details of CDEF}

\subsection{CDEF vs. the SPONGE vs. other methods}

\begin{figure}
\includegraphics[width=\linewidth]{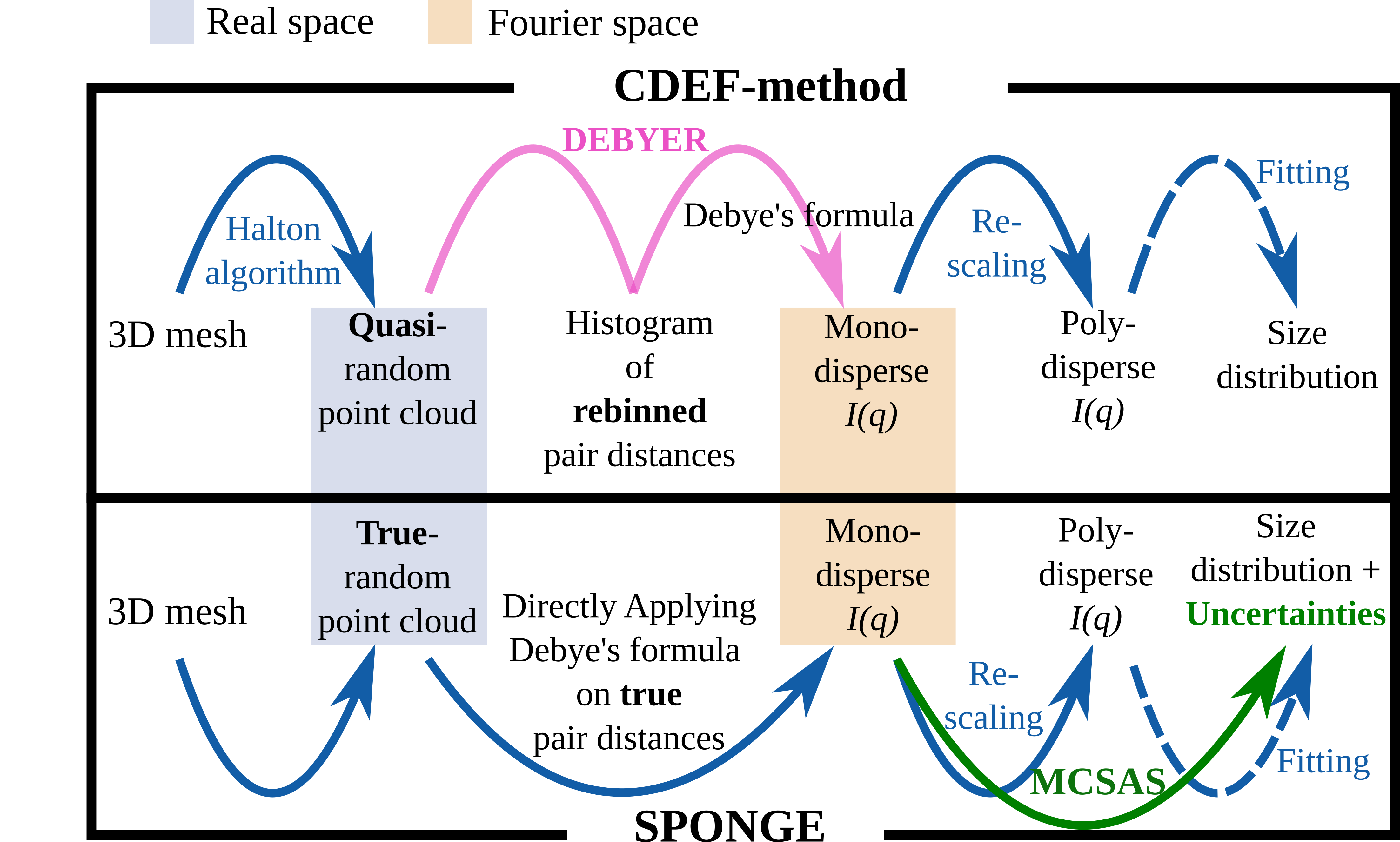}
\caption{Comparison between CDEF and the SPONGE. Both methods use Debye's scattering formula to calculate the single-particle scattering profile starting from the associated three-dimensional (3D) mesh which represents the arbitrary particle shape.}
\label{fig:debyer_vs_sponge}
\end{figure}

In this section, CDEF is compared to similar pre-existing approaches for the computation of form factors using Debye's scattering formula. 

The SPONGE has already proven to successfully simulate helicoidal supramolecular copolymers at different structural parameters \cite{aratsu_2020}, therefore it is used here to verify the accuracy of the much faster CDEF in order to use the latter with trust. In contrast to the SPONGE, which directly evaluates Debye's formula for each $q$, Debyer reduces the computational effort by a two-step process where the binned distance distribution histogram is first computed and subsequently reused for each value of $q$. This approximation should have a negligible effect if the bin size of the histogram is sufficiently small compared to the particle size.


By default, CDEF also employs a quasi-random filling as opposed to the true random filling as used by the SPONGE. As a result, a smaller number of punctiform scatterers can be used w.r.t. the same usable $q$-range which also reduces computational effort. 

All in all, the basic concept of CDEF originates from \cite{hansen_1990} as well as \cite{Pedersen_2012}. For instance, Hansen calculated $I(q)$ of a complex-shaped fibrinogen using the distance distribution function of a corresponding quasi-random point cloud \cite{hansen_1990}, without rebinning the distance distribution however. Moreover, Hansen focused on calculating single-particle $I(q)$, but did not consider polydisperse particle ensemble. Pedersen et al., on the other hand, evaluated polydisperse immune-stimulating-complex particles by applying Debye's equation. In doing so, the true-random particle cloud including $40 \, 000$ scatterers was split into $10$ subsets of $4 \, 000$ scatterers from which $10$ single pair distance histograms were calculated and then summed up to generate the total histogram \cite{Pedersen_2012}. Compared to the direct computation of the total histogram, this procedure lowers computational cost by one order of magnitude \cite{Pedersen_2012}, but at the expense of accuracy. 

Even though the approach of \citeasnoun{Pedersen_2012} is very similar to CDEF, corresponding software has not been published to our knowledge such that it would be accessible to SAXS experimentalists. 

Since CDEF and SPONGE use a random filling method (i.e., quasi-random or true random) instead of the exact atomic positions of the periodic lattice of particles, the computational time for both methods is significantly less than for a periodic lattice. For the evaluation of Au nanocubes with face-to-face distance $50 \, \mathrm{nm}$, for instance, choosing $30 \, 000$ rather than $\sim 5.2 \cdot 10^{6}$ scatterers which is the approximate number of atoms in a gold crystallite of this size \cite{pauling_1947}, decreases computational effort by four orders of magnitude ($\sim 30 \, 500$). The computation of a single SAXS profile using CDEF with $30 \, 000$ scatterers requires  $\sim 500 \, \mathrm{ms}$ on a single modern desktop computer with a quad core processor. Hence, this would need more than 4 hours in case of $5.2 \cdot 10^6$ scatterers, which justifies the simplification using the quasi-random filling instead of the exact crystallographic positions. 

In addition, CDEF allows to individually adjust the scattering length of each single scatterer as described by \citeasnoun{Pedersen_2012} leading to much more versatile applications such as fitting core-shell-structured particles.

\begin{figure}
\includegraphics[width=\linewidth]{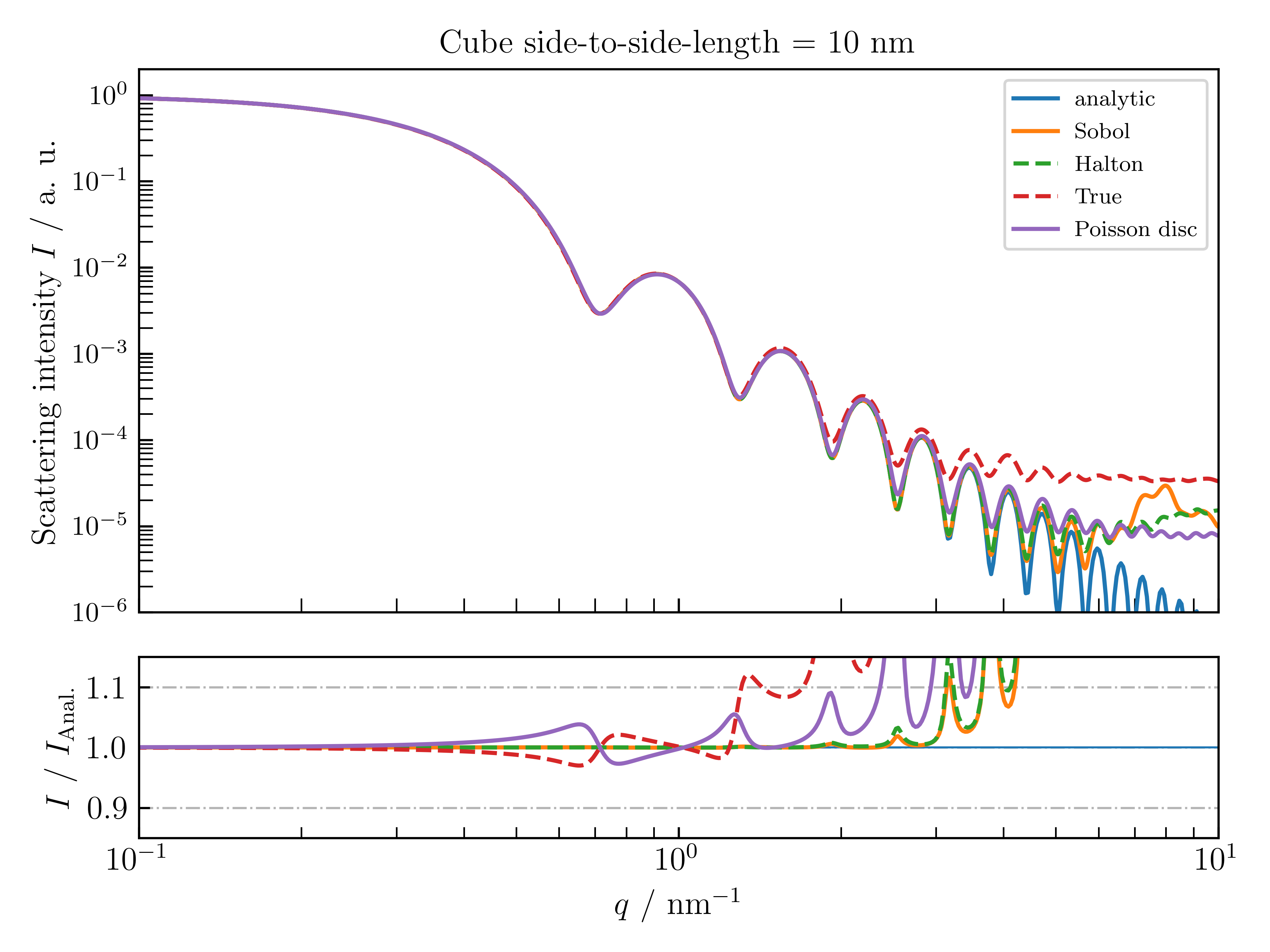}
\caption{Comparison of normalized single-particle SAXS profiles using CDEF with the exact analytic SAXS profile $I_{\mathrm{Anal.}}$ of a cube with edge length $=10 \, \mathrm{nm}$ and electron contrast $\Delta \rho = 1 \, \mathrm{nm}^{-3}$. All point clouds consist of $\sim 30 \, 000$ scatterers. Using a Poisson-disc-distribution does not result in a higher agreement of the corresponding scattering profile with the analytic profile when compared to the Sobol or Halton distribution.}
\label{fig:debyer_result_0}
\end{figure}

Besides the use of a periodic grid for particle formation, the use of a Poisson disc algorithm would also be conceivable, but it converges slower and slower with increasing scattering point number than the use of a random distribution method. Figure \ref{fig:debyer_result_0} shows here that the Poisson disc method is not closer to the analytical solution.

For SAXS on isotropic macromolecules, similar approaches have been made in the past by the European Molecular Biology Laboratory (EMBL) to calculate $I(q)$ from the scattering properties of the underlying substructure by introducing CRYSOL, which are part of the ATSAS software package \cite{atsas_2017, Crysol_1995}.
However, unlike CDEF, CRYSOL calculates $I(q)$ of macromolecules in dilute solution from the \textit{exact} crystallographic positions and form factors of their individual atoms \cite{Crysol_1995}. This involves spherical averaging using spherical harmonics and their orthogonality properties to obtain a simplified expression for the total molecular form factor.
Since CRYSOL also accounts for scattering from the missing water molecules and the hydration shell of the molecules as correction terms for the total form factor of the molecules, it can resolve scattering profiles up to $q \leq 4 \, \mathrm{nm}^{-1}$ \cite{Crysol_1995}. 
However, despite the higher $q$ range, it is not necessary to use accurate crystallographic positions to adequately calculate $I(q)$ of suspended nanoparticles. Moreover, the electron contrast of particles composed of metals or oxides is much higher than that of biomolecules, which allows us to ignore a possible hydration shell for accurate SAXS data evaluation.
Another software from ATSAS called DAMMIN calculates scattering patterns based on a substructure consisting of artificial scatterers called dummy atoms, which is generally more similar to CDEF than CRYSOL \cite{Dammin_1999, atsas_2017}. 

Lastly, there is the comprehensive software package DEBUSSY using Debye's formula 
which offers users a GUI and therefore allows one to compute scattering pattern of crystalline or non-ordered nanostructures without any programming knowledge \cite{debussy_2015}. Even though users of DEBUSSY are able to generate nano-scale clusters, such as nanoparticles or quantum dots, by defining the coordinates of the underlying (atomic) structure to calculate the corresponding SAXS scattering profiles \cite{bertolotti_2016}, CDEF overall seems to be more suited to analyze SAXS scattering pattern of nanoparticles in the upper nano-scale region due to the much easier cloud building feature of simply loading an stl-file defining the particle's shape and then choosing the desired filling algorithm.

\newpage
\section{CDEF vs. analytic formulae}

This section is intended as a supplement to the comparison between numerically calculated and analytical form factors. Here we also present background modeled scattering curves by changing the number of filled bins of the corresponding pair distance distribution. 

For the true random distribution, it is generally recommended to exclude only the self-correlation as suggested by \citeasnoun{Pedersen_2012}, which correpsonds to the first histogram bin with zero distance. For the quasi-random case, users of CDEF have several options. According to the comparisons shown below, the best correspondence in the mid $q$ range is achieved by not deleting any bins. However, if this leads to an excessive background for higher $q$ values and the number of scatterers cannot be increased, a few histogram bins can be deleted, starting with the second one, and leave the first bin untouched. In this chapter, we show examples for the different particle shapes of how many consecutive bins need to be deleted for the curves to be closest to the analytical solution.

\begin{figure}
\includegraphics[width=\linewidth]{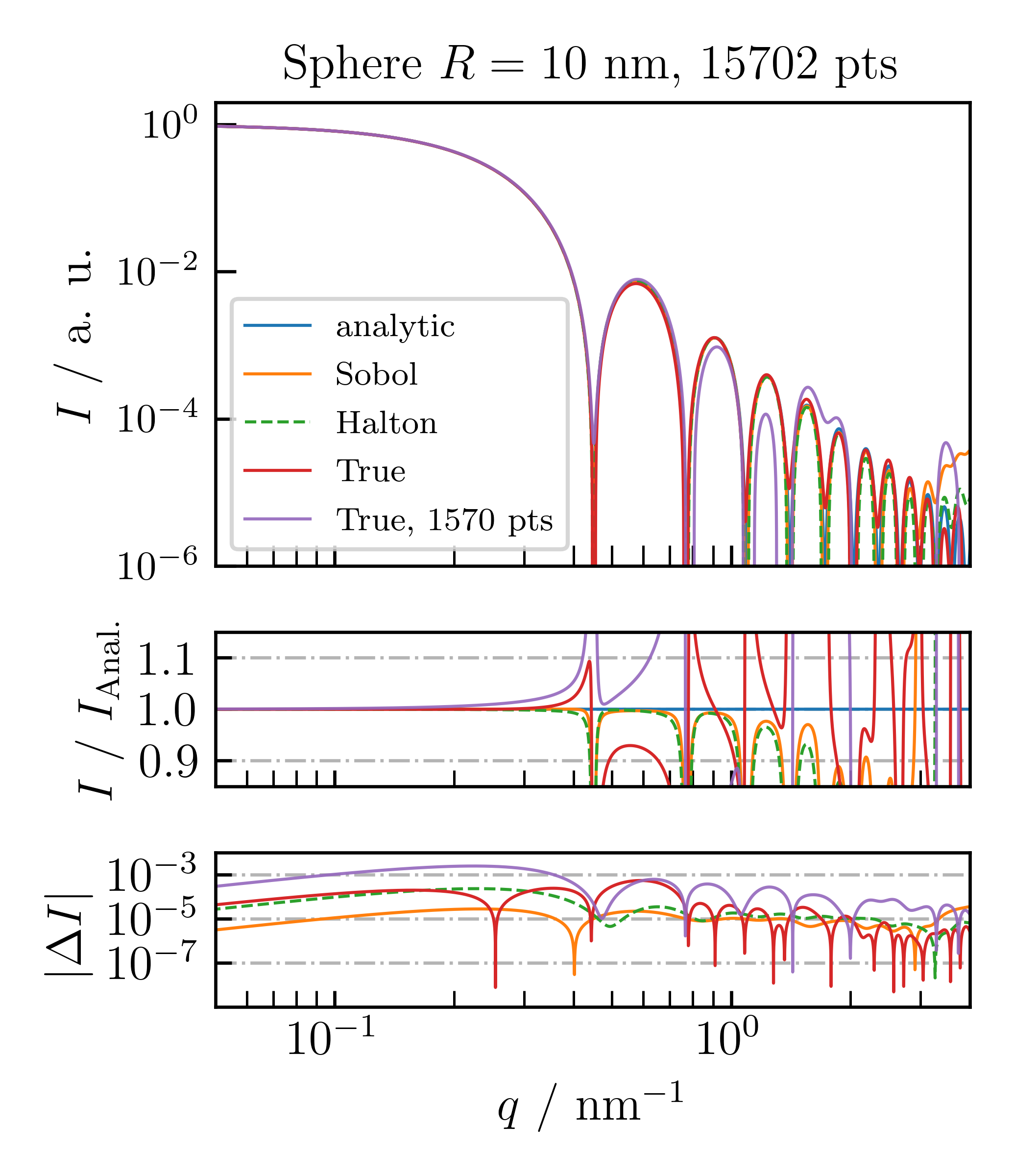}
\caption{Comparison of normalized single-particle SAXS profiles using CDEF with the exact analytic SAXS profile $I_{\mathrm{Anal.}}$ of a sphere with radius $R=10 \, \mathrm{nm}$ and electron contrast $\Delta \rho = 1 \, \mathrm{nm}^{-3}$. Manual changes to the pair distance histogram allow user of CDEF to increase the usable $q$-range of the numeric curves through \textit{modeling of the artificial background signal}.}
\label{fig:debyer_result_1}
\end{figure}

\begin{figure}
\includegraphics[width=\linewidth]{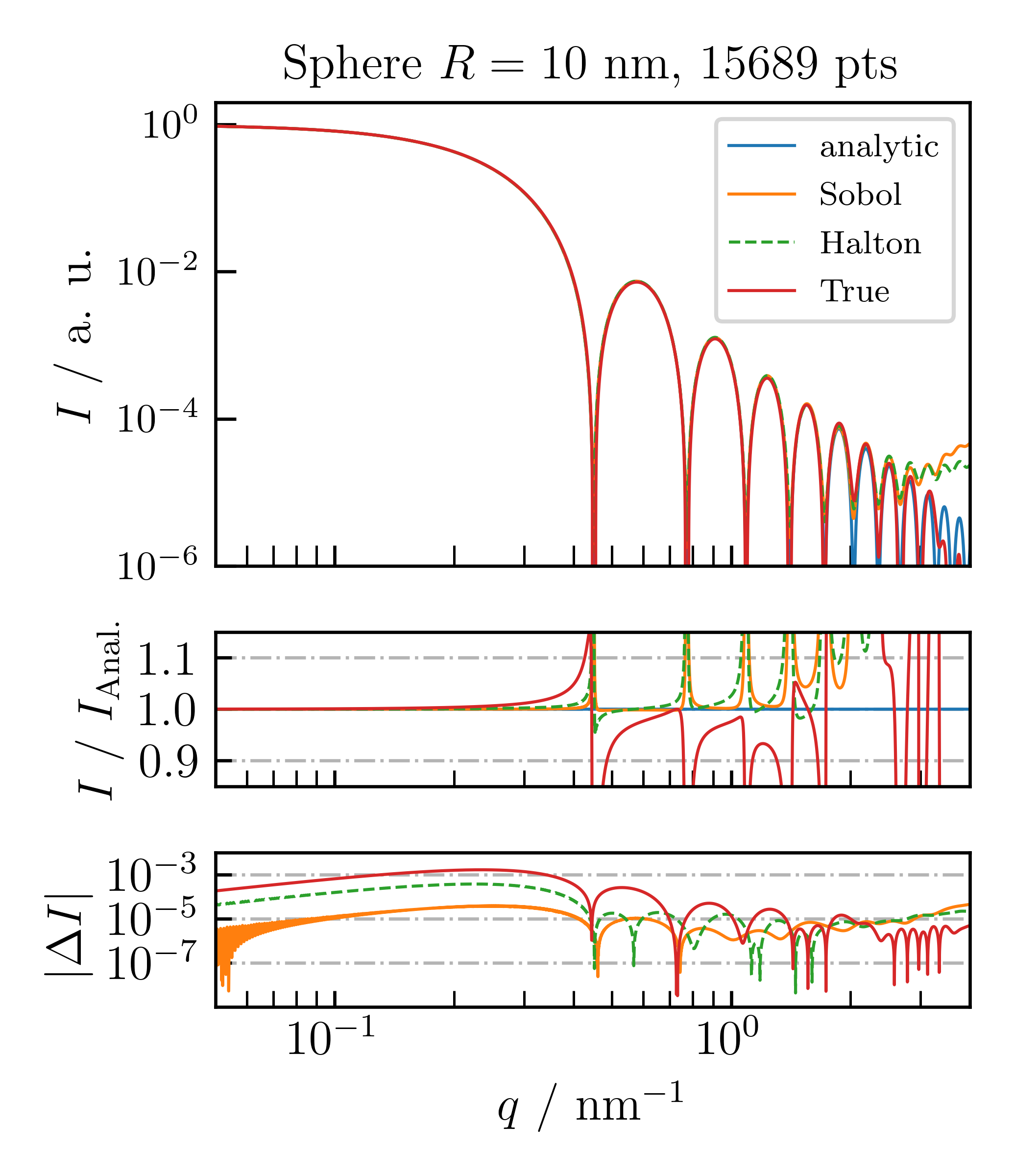}
\caption{Comparison of normalized single-particle SAXS profiles using CDEF with the exact analytic SAXS profile $I_{\mathrm{Anal.}}$ of a sphere with radius $R=10 \, \mathrm{nm}$ and electron contrast $\Delta \rho = 1 \, \mathrm{nm}^{-3}$. Only the true-random profile is background modeled ($n_{k=0}=0$).}
\label{fig:debyer_result_2}
\end{figure}

For proper modeling of the artificial background signal of the quasi-random SAXS pattern of the spherical cloud, the first $60$ of $10 \, 000$ bins were deleted ($n_{k=1, ... ,60} = 0$), except the first bin $n_{k=0}$ to avoid excessive background modeling. The background modeled SAXS profiles of the quasi- and true-random ($n_{k=0}=0$) filling algorithms are quite similar with a small advantage for the quasi-random distributions in the lower and middle $q$-region w. r. t. ($I \, \mathrm{/} \, I_{\mathrm{Anal.}}$)-plot (fig. \ref{fig:debyer_result_1}). Interestingly, this remains the case even if the background modeled true-random curve is compared to the non-modeled quasi-random profiles (fig. \ref{fig:debyer_result_2}).

\newpage

\begin{figure}
\includegraphics[width=\linewidth]{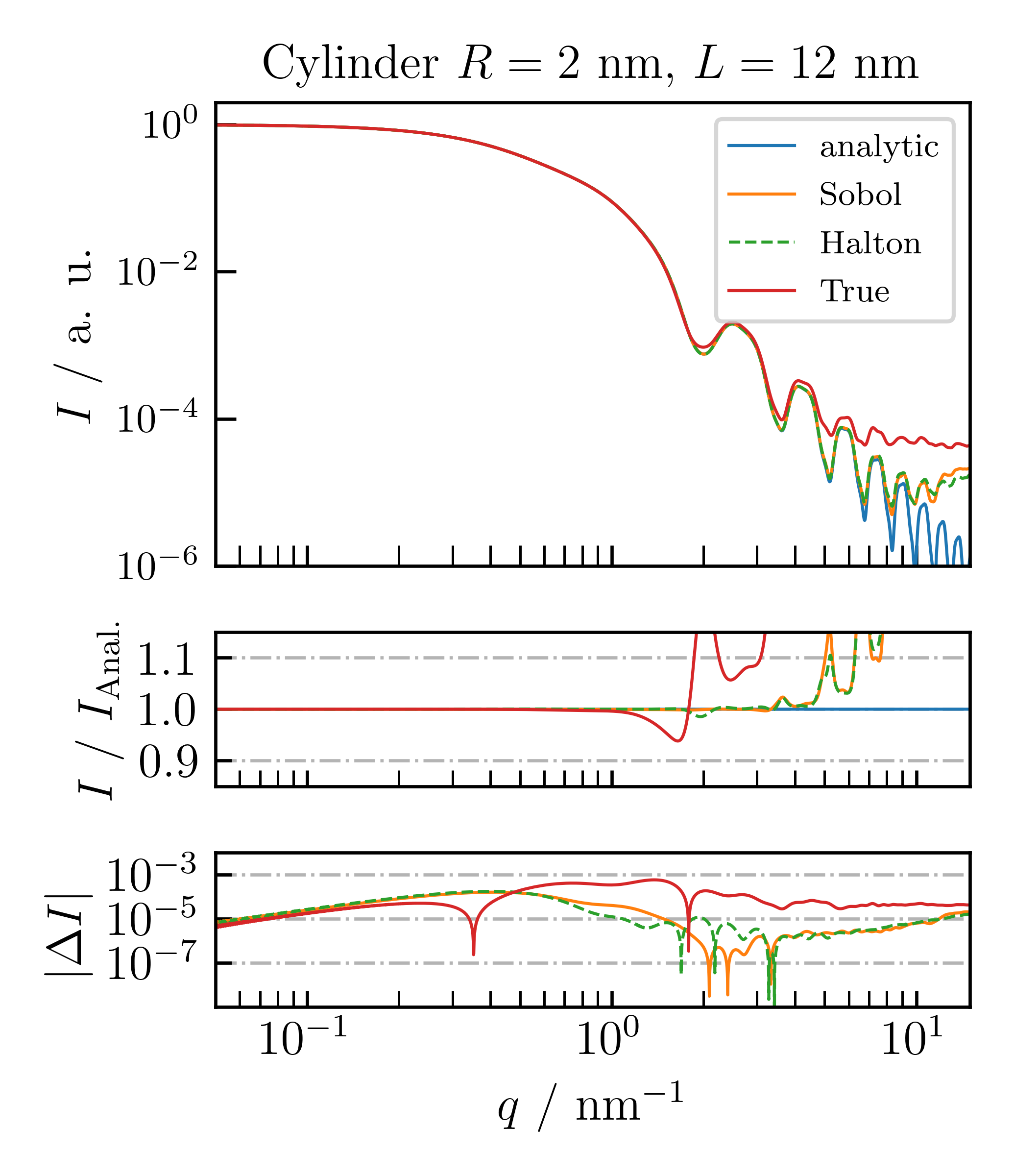}
\caption{Comparison between numeric and analytic single-particle SAXS profiles of a cylindrical particle with radius $R = 2 \, \mathrm{nm}$, length $L= 12 \, \mathrm{nm}$ and electron contrast $\Delta \rho = 1 \, \mathrm{nm}^{-3}$.}
\label{fig:debyer_result_3_2}
\end{figure}

In order to evaluate the performance of CDEF for particles with lower symmetry, fig. \ref{fig:debyer_result_3_2} compares analytically and numerically obtained single-particle SAXS profiles of a cylinder with aspect ratio $L \, \mathrm{/} \, R = 6$ and $\Delta \rho = 1 \, \mathrm{nm}^{-3}$. The analytic SAXS curves were calculated using the well-known expression \cite{guinier_1955, Galantini_2004}.
For the calculation of each numeric profile, cylindrical clouds of $N \approx 30 \, 000 \, \pi \mathrm{/} 4 \approx 23 \, 560$ scattering points were generated similarly to the spherical clouds before. The quasi-random profiles match the analytic one up to the $4$th local maximum, whereas the true-random pattern hardly matches the analytic solution up to the first local maximum. 

\begin{figure}
\includegraphics[width=\linewidth]{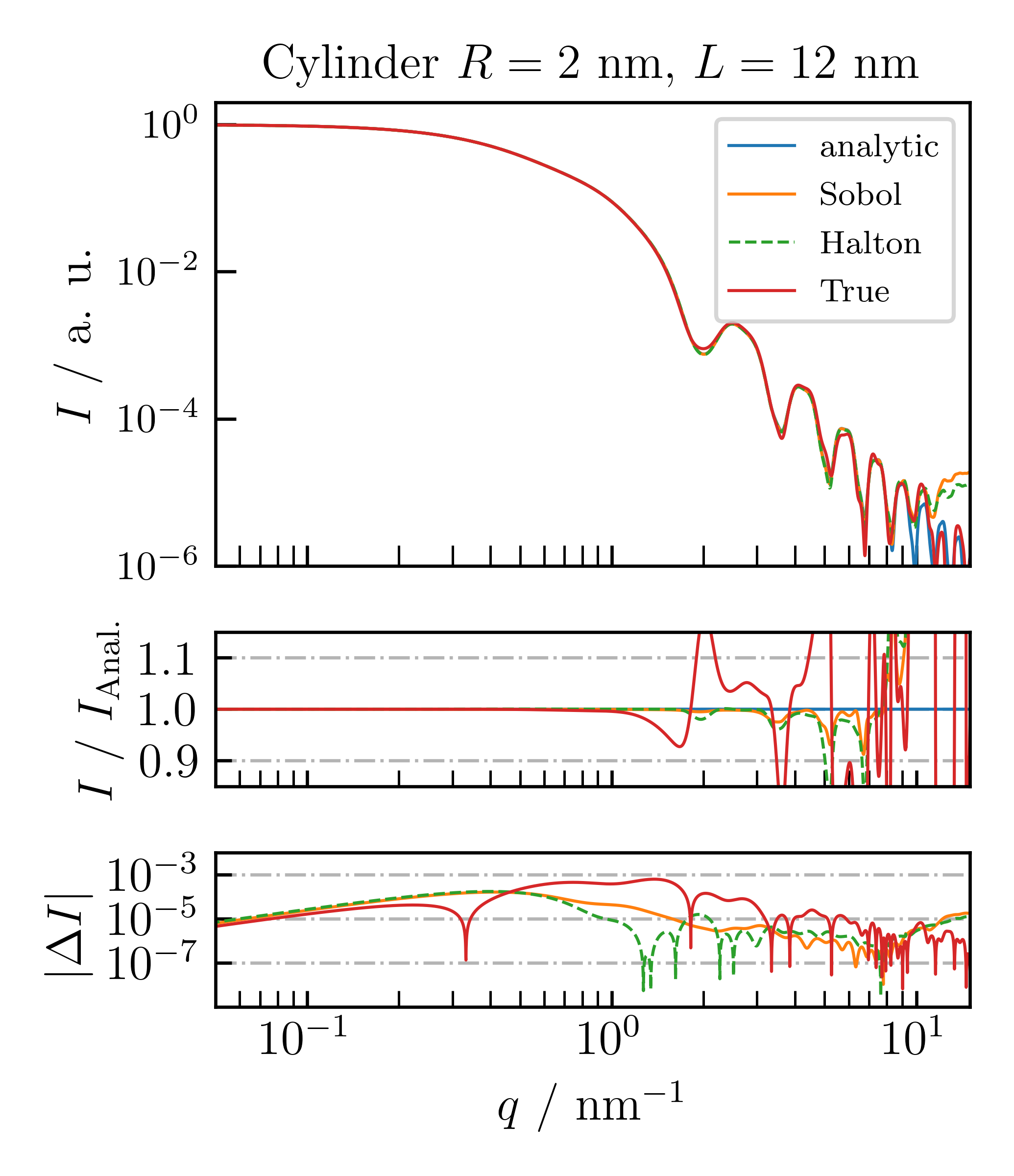}
\caption{Comparison between numeric and analytic single-particle SAXS profiles of a cylindrical particle with radius $R = 2 \, \mathrm{nm}$, length $L= 12 \, \mathrm{nm}$. All profiles are background modeled.}
\label{fig:debyer_result_3_2_35}
\end{figure}

After deletion of the first bin ($n_{k=0} = 0$) of the true-random histogram as well as $35$ out of $10 \, 000$ ($n_{k=1, ... ,35} = 0$) bins of the quasi-random histogram, the quasi-random SAXS profiles are still more consistent with the analytic solution regarding the lower and middle $q$-range (fig. \ref{fig:debyer_result_3_2_35}). The specific upper bin limit of $35$ for the cylinder was determined by gradually clearing bins until an adequate background modeling was obtained, as done for the spherical and the cubic point cloud. As for the spherical cloud, comparing the background modeled true-random curve to the non-modeled quasi-random profiles indicates that the quasi-random distribution seems to be superior in this $q$-region (fig. \ref{fig:debyer_result_3_2} vs. \ref{fig:debyer_result_3_2_35}).

\newpage

\begin{figure}
\includegraphics[width=\linewidth]{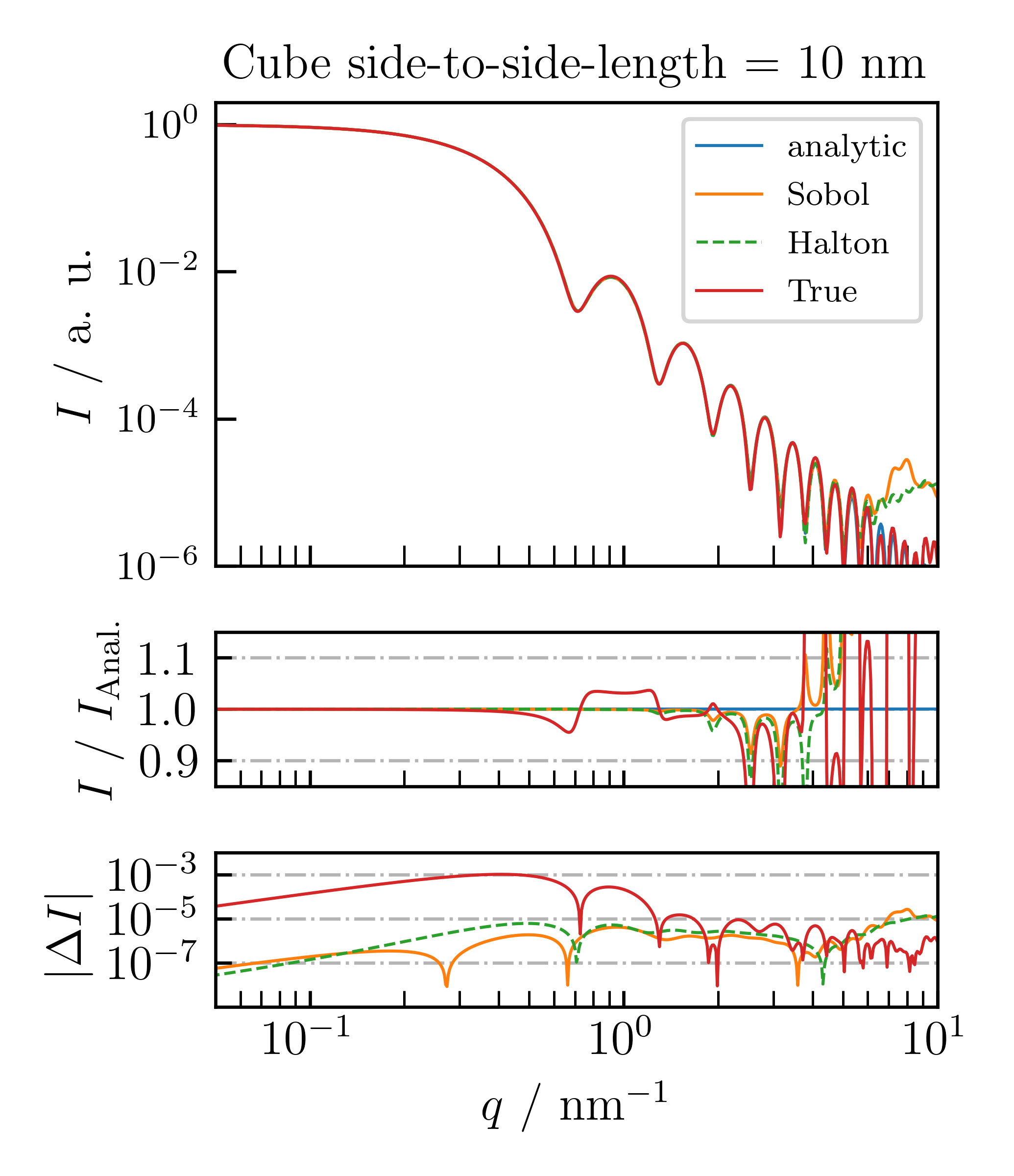}
\caption{Comparison between numeric and analytic background modeled SAXS profiles of an ideal cube with side-to-side length $10 \, \mathrm{nm}$ and electron contrast $\Delta \rho = 1 \, \mathrm{nm}^{-3}$. In the lower and middle $q$-region, the quasi-random profiles are more consistent with the analytic profile compared the true-random one.}
\label{fig:debyer_result_cube}
\end{figure}

Regarding ideal cubes, CDEF is also able to adequately match $I_{\mathrm{Anal.}}$ as depicted in fig. \ref{fig:debyer_result_cube}. The analytic curve was calculated by averaging the expression from \cite{mittelbach1961rontgenkleinwinkelstreuung} over two independent spatial coordinates. For this particular cube with side-to-side length $10 \, \mathrm{nm}$, CDEF with $30 \, 000$ scatterers as well as $I_{\mathrm{Anal.}}$ were evaluated on $999$ $q$-values, whereas for the calculation of each numeric profile CDEF needed only $\sim 500 \, \mathrm{ms}$.
Background modeling was conducted by clearing the first $45$ of $10 \, 000$ bins ($n_{k=1, ... ,45} = 0 $) for the quasi-random profiles as well as the first bin ($n_{k=0} = 0$) for the true-random profile. This specific upper bin limit of $45$ for the cube was determined by gradually clearing bins until an adequate background modeling was obtained, as done for the spherical and the cylindrical point cloud.

\newpage
\section{Diverse models of cubic particles}
This section describes the construction of the cubic shape with truncated edges. 
\subsection{Hessian normal - cube with truncated edges}
The Hessian normal form was used to parameterize the degree of truncation with respect to the cubic model with symmetrically cut edges. It describes the shortest distance $D$ of a point with the Euclidean position $\vec{x}$ relative to a given plane described by a support vector $\vec{a}$ and normal vector $\vec{n}$: 

\begin{eqnarray}
D = (\vec{x} - \vec{a}) \cdot \vec{n}.
\label{eq:intensity11}
\end{eqnarray}

For each scattering point with position $\vec{x}$, $D$ is calculated for each of the 12 sectional planes which cut the edges of the cubic model. If $D < 0$ for all 12 sectional planes, the point is located inside the cube, because the defined normal vectors $\vec{n}$ are pointing away from the cloud's center by definition. All outside points are deleted by setting their corresponding form factor to zero. The degree of truncation can be influenced by equally varying the length of all $\vec{a}$ by modifying an introduced truncation factor $T$, with

\begin{eqnarray}
\vec{a} = T \frac{L}{\sqrt{2}} \, \vec{n}.
\label{eq:intensity12}
\end{eqnarray}

With this definition, $T = 1$ indicates no truncation and corresponds to the \textit{ideal} cube.

\newpage
\section{Results and Discussion}
With this chapter we want to complement the chapter of the published work, where only the results of the cubic model with rounded edges are presented. In addition, this chapter contains the fits with an ideal cube as well as a cube with truncated edges. First, the results for the ideal cube are shown, followed by the truncated cube. 

\begin{figure}
\includegraphics[]{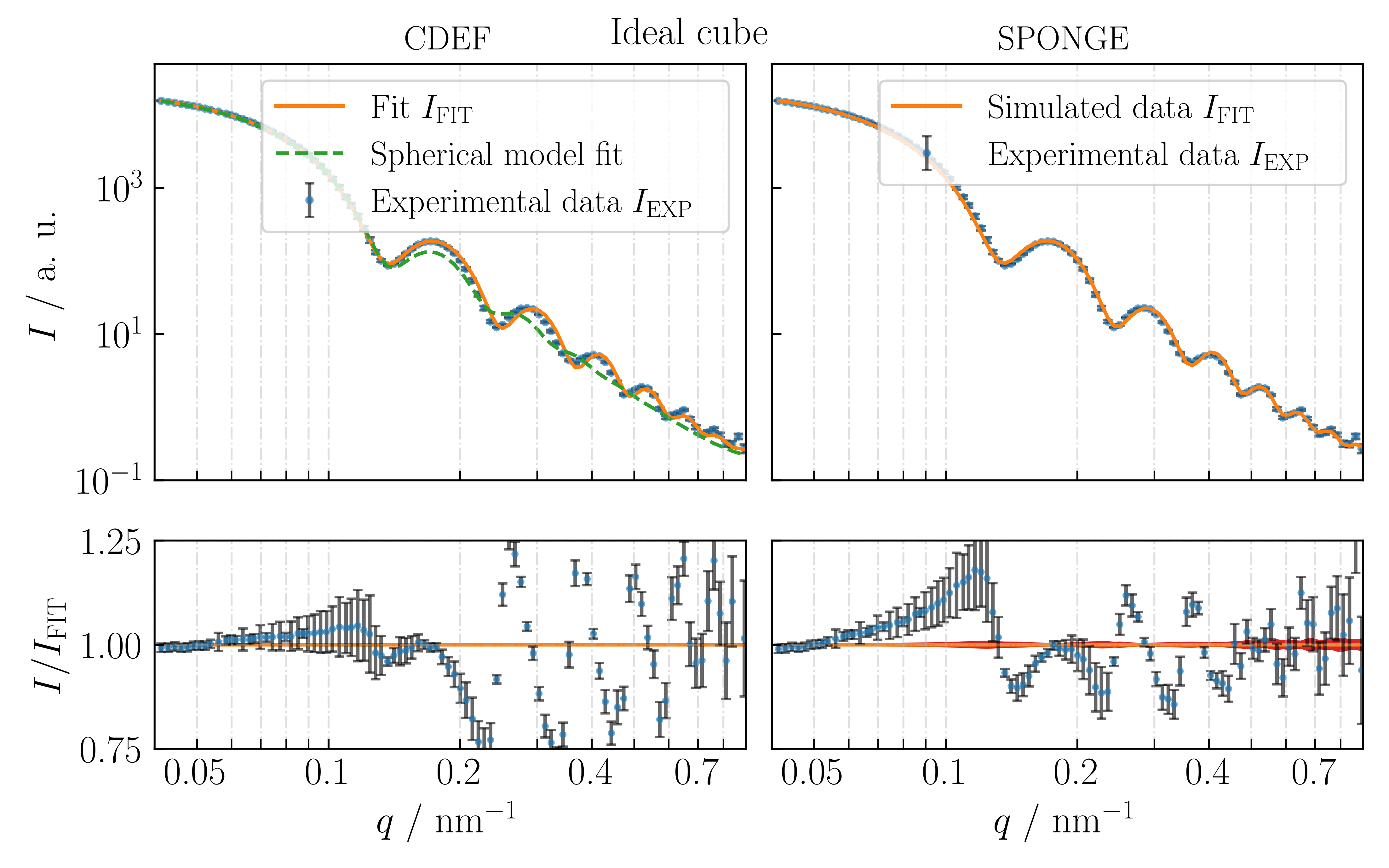}
\caption{CDEF versus the SPONGE. Fit results of Au nanocubes using the model of an ideal cube. Coupling of the SPONGE with MCSAS additionally reveals an uncertainty of $I_{\mathrm{FIT}}$ marked in red the SPONGE's $I \mathrm{/} I_{\mathrm{FIT}}$-Plot, thus an uncertainty of the underlying size distribution can be stated (fig. \ref{fig:hist_ideal}).}
\label{fig:saxs_ideal}
\end{figure}

For comparison, a spherical model was additionally included in the evaluation as depicted in fig. \ref{fig:saxs_ideal}. The weak match between experimental data and spherical fit demonstrates the sensitivity of the scattering experiment to the particle shape. Even though the ideal cubic model also leads to pronounced deviations at $0.2 \, \mathrm{nm}^{-1} < q$, in general it matches the oscillations with better agreement compared to the spherical model which is also confirmed by the corresponding values of $\chi^2$. The spherical model only matches the Guinier region satisfyingly resulting in a mean diameter of $d = 65.5 \, \mathrm{nm}$ with a standard deviation of $\sigma_{\mathrm{d}} = 6.7 \, \mathrm{nm}$.

Using the ideal cubic model reveals a volume-weighted mean face-to-face-distance of $L = 52.5 \, \mathrm{nm}$ with a distribution width of $\sigma_{\mathrm{L}} = 2.8 \, \mathrm{nm}$. With the SPONGE we get an mean value of $L = (53.20 \pm 0.06) \, \mathrm{nm}$ with a distribution width of $\sigma_{\mathrm{L}} = (2.7 \pm 1.0) \, \mathrm{nm}$ which leads to a relative deviation of $\Delta L \, \mathrm{/} \, L \approx 1.5\,\%$ between CDEF and the SPONGE. Regarding the Guinier region, CDEF seems to give a slightly better fit compared to the SPONGE, whereas in the Porod region the SPONGE is superior.

\begin{figure}
\includegraphics[]{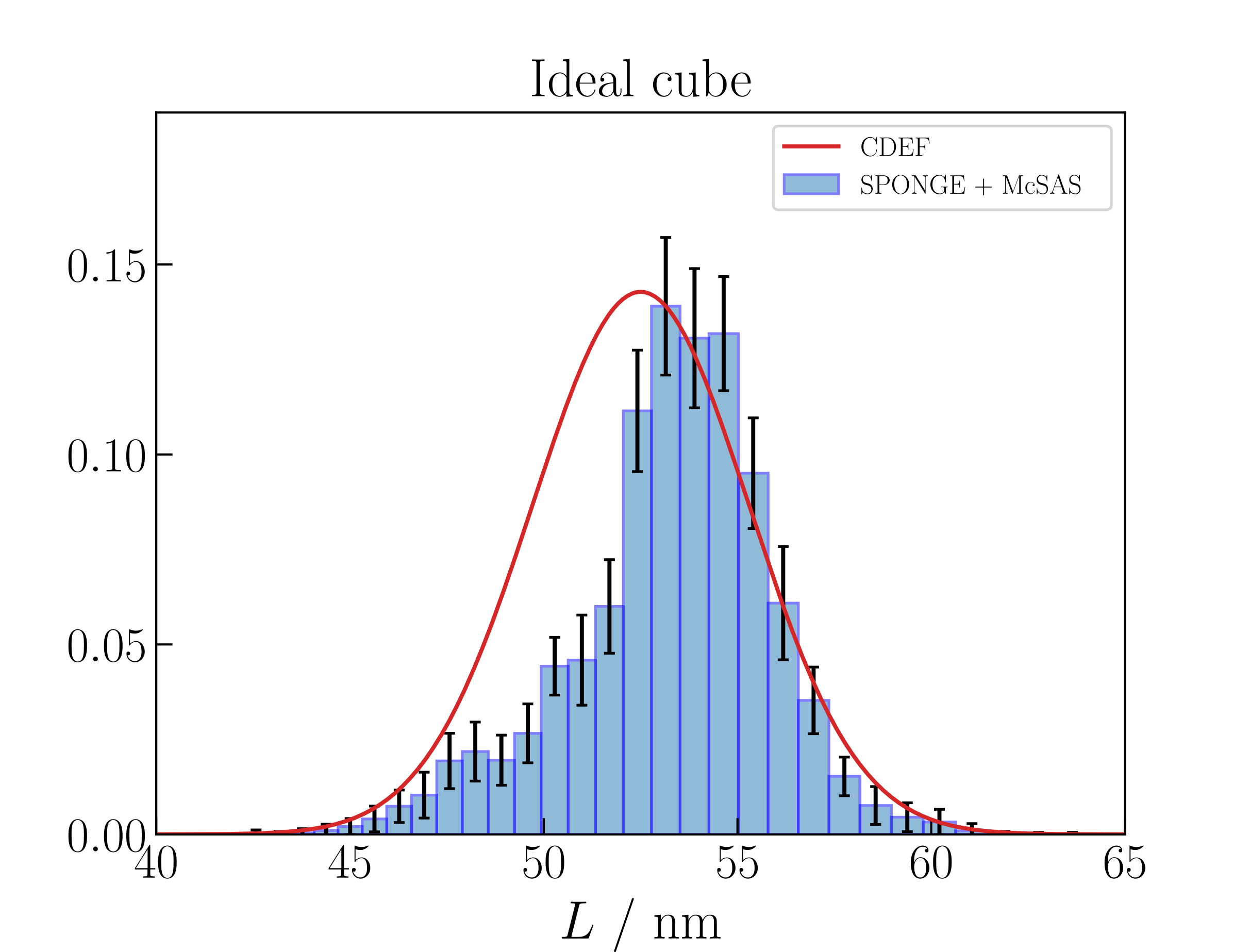}
\caption{CDEF vs. the SPONGE: Normalized distribution of side-to-side length of the Au nanocubes. The uncertainty of the volume-weighted distribution using the SPONGE with a mean value of $L = (53.20 \pm 0.06) \, \mathrm{nm}$ is indicated by error bars. The volume-weighted distribution using CDEF reveals a mean value of $L = 52.5 \, \mathrm{nm}$.}
\label{fig:hist_ideal}
\end{figure}

The truncated cubic model reveals a mean value of $L = 53.4 \, \mathrm{nm}$ with $\sigma_{\mathrm{L}} = 3.3 \, \mathrm{nm}$, and a truncation factor of $T = 0.91$. From a geometrical perspective, this truncation factor means that on average $\sim 2.6 \, \mathrm{nm}$ are cut off on both sides of each edge. With the SPONGE we obtain a value of $L = (54.00 \pm 0.04) \, \mathrm{nm}$ with a distribution width of $\sigma_{\mathrm{L}} = (3.0 \pm 0.8) \, \mathrm{nm}$ is obtained. A reason for the relative deviation of $\Delta L \, \mathrm{/} \, L \approx 1.1\,\%$ between CDEF and the SPONGE (as for the other cubic models) may result from the fact that CDEF is confined to a Gaussian size distribution, whereas the SPONGE is not. This assumption is also confirmed by the fact that there are no deviations in the corresponding single-particle and polydisperse scattering profiles when comparing CDEF with the SPONGE.

\begin{figure}
\includegraphics[]{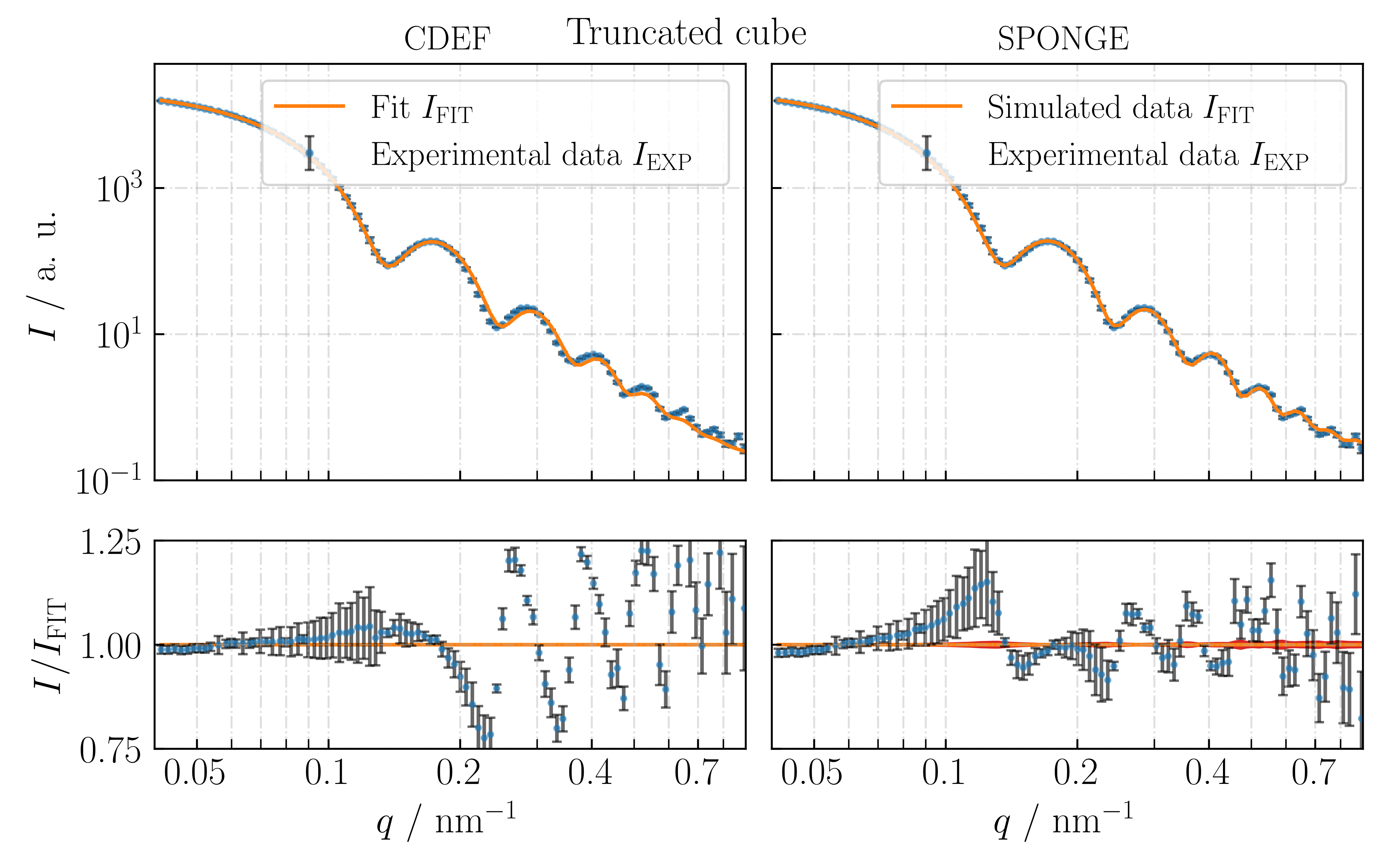}
\caption{CDEF versus the SPONGE. Fit results of Au nanocubes using a cubic model with truncated edges. Coupling of the SPONGE with MCSAS additionally reveals an uncertainty of $I_{\mathrm{FIT}}$ marked in red as explained in fig. \ref{fig:saxs_ideal}, thus an uncertainty of the underlying size distribution can be stated (fig. \ref{fig:hist_truncated}).}
\label{fig:saxs_truncated} 
\end{figure}

\begin{figure}
\includegraphics[]{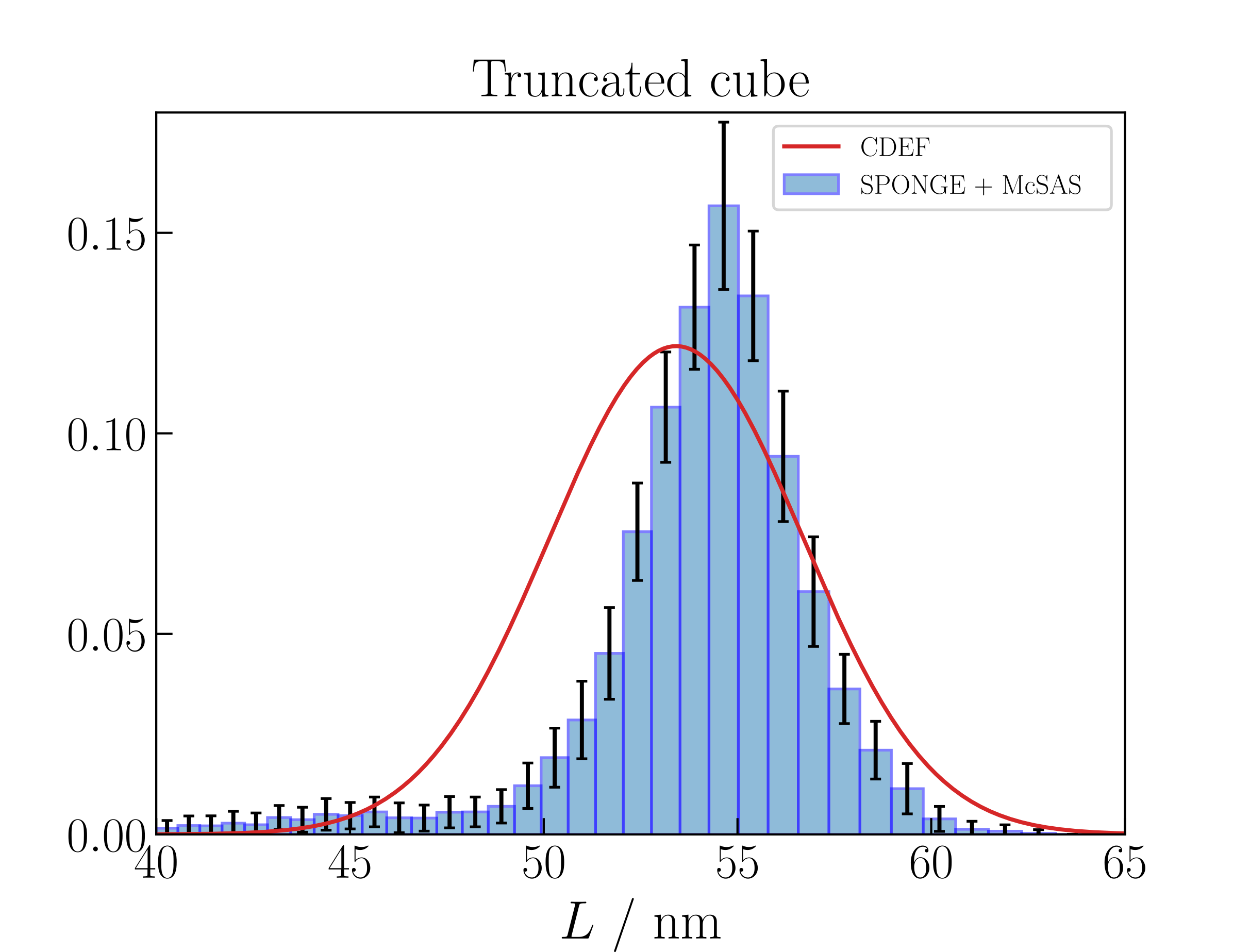}
\caption{CDEF vs. the SPONGE: the SPONGE's volume-weighted size distribution reveals a mean value of $L = (54.00 \pm 0.04) \, \mathrm{nm}$. The volume-weighted distribution using CDEF shows an expectation value of $L = 53.4 \, \mathrm{nm}$.}
\label{fig:hist_truncated}
\end{figure}

\referencelist{supp}